\documentclass[a4paper,11pt]{article}
\usepackage[top=2.54cm,bottom=2.54cm, left=2.54cm, right=2.54cm]{geometry}
\usepackage{amsmath}
\usepackage{amsfonts}
\usepackage{amssymb}
\usepackage{graphicx}
\usepackage{bm}
\usepackage{enumerate}
\usepackage{color}
\usepackage[sort,comma]{natbib}
\usepackage{float}
\usepackage[colorlinks=true, allcolors=blue]{hyperref}
\usepackage{changepage}
\usepackage{dirtytalk}
\usepackage{pdflscape}
\usepackage{booktabs}
\usepackage{comment}

\usepackage{amsthm}
\newtheorem{theorem}{Theorem}[section]

\theoremstyle{definition}

\newtheorem{assumption}{Assumption}[section]

\theoremstyle{remark}
\newtheorem{remark}{Remark}[section]

\newcommand{\thetao}{\theta_0}
\newcommand{\thetahat}{\hat{\theta}_n}
\newcommand{\convp}{\xrightarrow{p}}

\newcommand{\R}{\mathbb{R}}
\newcommand{\E}{\mathbb{E}}
\newcommand{\norm}[1]{\left\|#1\right\|}
\newcommand{\convd}{\xrightarrow{d}}
\newcommand{\OmegaMat}{\Omega(\thetao)}
\newcommand{\GMat}{G(\thetao)}

\makeatletter
\def\vec{\mathop{\operator@font vec}\nolimits}
\makeatother

\DeclareMathOperator{\diag}{diag}

\title{The $\alpha$--regression for compositional data: a unified framework for standard, temporal and spatial regression models including compositional predictors}
\author{Michail Tsagris$^1$, Nader Alharbi$^2$, Abdulaziz Alenazi$^3$ and Yannis Pantazis$^4$\\
\\
$^1$ Department of Economics, University of Crete, Greece  \\
\href{mailto:mtsagris@uoc.gr}{mtsagris@uoc.gr} \\
$^2$  King Saud bin Abdulaziz University for Health Sciences, and \\
King Abdullah International Medical Research Center, Riyadh, Saudi Arabia \\
\href{mailto:alharbina@ksau-hs.edu.sa}{alharbina@ksau-hs.edu.sa} \\
$^3$ Department of Mathematics, College of Science \& Center for \\
Scientific Research and Entrepreneurship, Northern Border University, Arar, Saudi Arabia, \\ \href{mailto:a.alenazi@nbu.edu.sa}{a.alenazi@nbu.edu.sa} \\
$^4$ Institute of Applied \& Computational Mathematics, \\
Foundation for Research \& Technology--Hellas,
Heraklion, Greece \\
\href{mailto:pantazis@iacm.forth.gr}{pantazis@iacm.forth.gr}
}

\begin{document}

\maketitle

\begin{center}
\textbf{Abstract}
\end{center}
We revisit the $\alpha$--regression framework for compositional data. We formulate $\alpha$--regression as a non--linear least squares problem, study its asymptotic properties, and provide efficient estimation via the Levenberg--Marquardt algorithm. We then propose a permutation--based hypothesis testing procedure, derive marginal effects for interpretation, and provide a visual inspection of the effect of each predictor. We further discuss robustified versions, the inclusion of natural splines, and the incorporation of compositional predictors, which further facilitate the formulation of a simple time series model. The framework is extended to spatial settings through four models. (a) The $\alpha$--spatially--lagged X regression model, which incorporates spatial spillover effects via spatially--lagged covariates, with decomposition into direct and indirect effects. (b) The $\alpha$--spatial autoregressive model that allows for spatial autocorrelation. (c) The geographically--weighted $\alpha$--regression, which allows coefficients to vary spatially for capturing local relationships. (d) The $\alpha$--eigenvector spatial filtering that is computationally efficient and captures spatial dependence via the eigenvectors of the kernelized distance matrix. Applications to four real datasets illustrate that the models perform on par with or outperform existing models in the literature. The examples showcase that $\alpha$--regression can outperform various competing regression models under different scenarios and its spatial extensions capture the dependence and improve the predictive performance. Overall, the examples provide evidence that the log--ratio methodology does not always lead to the optimal results. \\
\\
\textbf{Keywords}: compositional data, $\alpha$--transformation, spatial regression

\section{Introduction}
Compositional data are characterized as vectors of non--negative components constrained to sum to a constant, conventionally normalized to unity. These data structures arise across diverse scientific domains, as evidenced by the substantial body of methodological literature devoted to their rigorous statistical analysis\footnote{For an extensive compilation of domain-specific applications involving compositional data, see \citep{tsagris2020}.}, including most recent applications in psychology \citep{lehmann2024improving, brown2016thurstonian}. The sample space of such data is defined by the standard simplex
\begin{eqnarray} \label{simplex}
\mathbb{S}^{D-1}=\left\lbrace(y_1,...,y_D) \ \big | \ y_i \geq 0,\sum_{i=1}^{D}y_i=1\right\rbrace, 
\end{eqnarray}
where $D$ denotes the dimensionality of the compositional vector.

The methodological imperative to develop models specifically calibrated for compositional data has catalyzed considerable innovation, particularly in the contemporary statistical literature. The foundational framework was established by \cite{ait2003} and predicated upon log--ratio transformations, thereby inaugurating the \textit{log--ratio analysis} (LRA). This methodology was subsequently refined by \cite{ilr2003}, who implemented an isometric log--ratio (ilr) transformation to preserve geometric properties. In contrast, the \textit{stay--in--the--simplex approach} employs probability distributions and regression structures intrinsically defined on the simplex manifold. Notably, Dirichlet regression has been extensively utilized within compositional frameworks \cite{gueorguieva2008, hijazi2009, melo2009}. Furthermore, \cite{iyengar2002} investigated the generalized Liouville distribution family, which accommodates negative or heterogeneous correlation structures, thereby extending beyond the restrictive positive correlation constraint of Dirichlet distributions. A less theoretically justified, yet occasionally employed strategy involves disregarding the unit--sum constraint and treating compositional data within a Euclidean framework---an approach designated as \textit{raw data analysis} (RDA) \citep{baxter2001, baxter2005}. A fourth methodological paradigm employs transformations, such as the square root \citep{scealy2011, scealy2014} or more general power transformations. One such power transformation is the $\alpha$--transformation \citep{tsagris2011}, which interpolates continuously between the RDA and LRA, thereby affording enhanced model flexibility while accommodating zero components naturally.

Given the ubiquity of compositional data across scientific disciplines, regression frameworks incorporating covariates have become methodologically essential. Applications encompass glacial sediment compositions, household expenditure allocations, geochemical soil analyses, morphometric measurements, electoral outcomes, pollution indices, and energy consumption patterns, among numerous other domains. The literature documents extensive applications of compositional regression methodology. For instance, \cite{ait2003} analyzed foraminiferal compositions across bathymetric gradients in oceanographic research. In hydrochemistry, \cite{otero2005} employed regression techniques to discriminate anthropogenic from lithogenic sources of riverine contamination in Spain. Economic research, exemplified by \cite{morais2018}, has related market share dynamics to explanatory variables, while political scientists have modeled candidate vote proportions as functions of demographic and socioeconomic predictors \citep{katz1999}. Compositional methodologies have similarly been deployed in bioinformatic analyses of microbiome community structures \citep{xia2013, chen2016, shi2016}.

\cite{tsagris2015a} introduced a regression framework predicated upon minimization of the Jensen--Shannon divergence and \cite{alenazi2022} investigated the properties of the $\phi$--divergence regression models which are applicable to zero--inflated compositional data. 
A fundamental limitation of the aforementioned regression models concerns their inability to accommodate zero--valued components directly. Consequently, methodological developments have addressed this constraint through various approaches. \cite{scealy2011} proposed a transformation mapping compositional data to the unit hypersphere, introducing Kent regression which naturally accommodates structural zeros. \cite{tsagris2018} extended the Dirichlet regression paradigm to accommodate zero components, yielding zero--adjusted Dirichlet regression. Within econometric contexts, \cite{mullahy2015} formulated regression frameworks for fractional response\footnote{This is another term for compositional data met in the field of econometrics.} data exhibiting zero inflation. Additional econometric strategies suitable for zero--augmented compositional data are systematically reviewed in \cite{murteira2016}. 
Concerning spatial autocorrelation structures, the spatially--lagged X (SLX) model represents a parsimonious specification incorporating spatial dependence exclusively through exogenous covariates, thereby excluding spatial lags of the dependent variable \citep{lesage2009introduction, elhorst2014spatial}. The spatial autoregressive (SAR) model \citep{kazar2012spatial, shi2025high}, analogous to temporal autoregressive processes, posits that observations are influenced by proximate spatial neighbors. Specifically, the SAR model expresses the dependent variable as a function of both explanatory covariates and a spatially weighted average of neighboring dependent variable realizations. Geographically--weighted regression (GWR) constitutes a local regression methodology designed to capture spatially heterogeneous relationships \citep{brunsdon1996}. In contrast to conventional regression, which assumes parameter stationarity, GWR permits spatial nonstationarity through location--specific coefficient estimation. A modern approach is the eigenvector spatial filtering (ESF) \citep{griffith2019spatial}, which captures the spatial dependence via the eigenvectors of the kernelized distance matrix.

The integration of the spatial regression framework within compositional data analysis represents a relatively narrow research area\footnote{The literature in spatial modelling contains more studies, but most of them rely on log--ratio transformations prior to the application of standard spatial models.}. \cite{leininger2013} synthesized hierarchical Bayesian models for zero--inflated compositional data, incorporating spatial random effects to accommodate local variation. \cite{nguyen2021} and \cite{yoshida2021} developed a SAR specification, and GWR, respectively, both employing the ilr transformation for compositional responses. \cite{clarotto2022} introduced a novel power transformation, conceptually analogous to the $\alpha$--transformation, specifically calibrated for geostatistical modelling of compositional data. More recently, \cite{nguyen2026modeling} introduced the Dirichlet SAR model. The ESF has not yet been applied to a compositional data setting, to the best of our knowledge.

In this paper, we adopt a pragmatic methodological stance, particularly tailored to regression with compositional data. The principal contribution of this paper is a unified framework for regression modelling of compositional data, with and without spatial or temporal dependence, including compositional data on the predictor side. Our approach is applicable to, and generalizes, all regression models that have been developed for compositional data and they rely on the ilr transformation. 

We examine the $\alpha$--regression \citep{tsagris2015b}, that was proposed as a generalization of Aitchison's log--ratio regression \citep{ait2003}. The $\alpha$--regression offers several advantages. It can accommodate zero components without imputation, provides a continuum between power transformations and log-ratio methods, and allows the transformation parameter to be selected empirically. Further, its redictive performance is often higher compared to classical methods. A disadvantage is the reduced interpretability of regression coefficients compared to log--ratio approaches, but the use of marginal effects (MEs) and of the individual conditional expectation (ICE) plots \citep{goldstein2015} can overcome this obstacle and facilitate visualization of the effect of the predictor variables.  

First, we review the $\alpha$--regression model, examine it as a non--linear least squares minimization problem and use a, computationally efficient, modified Levenberg--Marquardt algorithm, to estimate the regression coefficients. We suggest two approaches to select the optimal value of $\alpha$, and then establish the consistency and the asymptotic normality of the regression coefficients. Concluding the presentation of the $\alpha$--regression, we discuss robust extensions and a simple method to incorporate compositional predictors, which allows for temporal associations. 

We next extend the $\alpha$--regression to accommodate spatial dependencies in four directions. The first extension is the $\alpha$--SLX model, where we allow for spatial correlation in the predictors, that is, we allow for spillover effects at the covariate level. The covariates affect directly the response, but also indirectly via the values of their neighbors. The second extension is the $\alpha$--SAR model, where we place the correlation on the response side. The response is not only affected by the covariates, but also by the response values of the neighbors. We then propose the GW$\alpha$R model, where the regression coefficients are location--specific. To alleviate the computational cost of the $\alpha$--SAR and GW$\alpha$R models, we propose the $\alpha$--ESF model. For all aforementioned spatial regression models, the selection of $\alpha$ and the free parameters is achieved via the spatial $K$--fold cross--validation (CV) protocol. 

The next section discusses the $\alpha$--regression, the choice of $\alpha$, the marginal effects, its asymptotic properties, and a simple method to incorporate compositional predictors. Section \ref{sec::spat} enhances the $\alpha$--regression to allow for spatial dependence by examining the four spatial extensions. Section \ref{sec:sims} presents a small simulaiton study regarding hypothesis testin. Section \ref{sec::app} illustrates the performance of all regression models on four real datasets, two of which contain spatial dependence. The first two datasets, when spatial dependence is ignored, are shown to be on par with or to outperform a non--linear model, the third dataset is a simple illustration showing that the $\alpha$--regression with compositional predictors outperforms a constrained regression model, whereas the fourth dataset shows that $\alpha$--regression can be used for time series, again outperforming a constrained regression model. Additionally, all spatial regression models are compared with another and some of their their weaknesses are revealed. Finally, Section \ref{sec::concl} concludes the paper. 

\section{The $\alpha$--regression}
First, the $\alpha$--transformation, used for the $\alpha$--regression, is defined, followed by the regression formulation.

\subsection{The $\alpha$--transformation}
For a composition $\bm{y} \in \mathbb{S}^{D-1}$, the centered log--ratio (clr) transformation is defined in Aitchison (1983) as 
\begin{eqnarray} \label{clr}
\bm{w}_0(\bm{y})=\left ( \log\left ({\frac{y_1}{\prod_{j=1}^Dy_j^{1/D}}} \right ),\ldots, \log\left ({\frac{y_D}{\prod_{j=1}^Dy_j^{1/D}}} \right ) \right ).
\end{eqnarray}
The sample space of (\ref{clr}) is the set
\begin{eqnarray} \label{Qd}
\mathbb{Q}_0^{D-1}=\left\lbrace \left(w_{1,0}, \ldots, w_{D,0} \right)^\top: \sum_{i=1}^Dw_{i,0}=0 \right\rbrace,
\end{eqnarray}
which is a subset of $\mathbb{R}^{D-1}$. Note that the zero sum constraint in (\ref{Qd}) is an obvious drawback of this transformation as it can lead to singularity issues. In order to remove the redundant dimension imposed by this constraint, one can apply the ilr transformation 
\begin{eqnarray} \label{ilr}
{\bf z}_0(\bm{y})=\bm{H}\bm{w}_0(\bm{y}),
\end{eqnarray}
where $\bm{z}_0(\bm{y})$ is a $D-1$ dimensional vector and $\bm{H}$ is the $(D-1) \times D$ Helmert \citep{helm1965} sub-matrix\footnote{This is the Helmert matrix after deletion of the first row. This sub--matrix is a standard orthogonal matrix in shape analysis used to overcome singularity problems. For further information, see \cite{dryden1998,le1999}.}. The sample space of Equation (\ref{ilr}) is $\mathbb{R}^{D-1}$ because left multiplication by the Helmert sub--matrix maps the clr transformed data  onto $\mathbb{R}^{D-1}$, thus, in effect, removing the zero sum constraint. 

\cite{tsagris2011} developed the $\alpha$--transformation as a more general transformation than the ilr (\ref{ilr}). Let 
\begin{eqnarray} 
\label{stayalpha}
{\bf u}_{\alpha}(\bm{y})=\left( \frac{y_1^{\alpha}}{\sum_{j=1}^Dy_j^{\alpha}}, \ldots, \frac{y_D^{\alpha}}{\sum_{j=1}^Dy_j^{\alpha}} \right)^\top
\end{eqnarray}
denote the power transformation for compositional data as defined by \cite{ait2003}, where $\alpha$ can take any value, but when zero values exist in the data, $\alpha$ can take only strictly positive values. In a manner analogous to Equations (\ref{clr}--\ref{ilr}), first define
\begin{eqnarray} \label{alef}
{\bf w}_{\alpha}(\bm{y})=\frac{D{\bf u}_{\alpha}(\bm{y})-1}{\alpha}.
\end{eqnarray}
The sample space of Equation (\ref{alef}) is 
\begin{eqnarray*}
\mathbb{Q}_{\alpha}^{D-1}=\left\lbrace \left(w_{1, \alpha}, \ldots, w_{D, \alpha} \right)^\top:\frac{-1}{\alpha} \leq w_{i, \alpha} \leq \frac{D-1}{\alpha},\sum_{i=1}^Dw_{i,\alpha}=0 \right\rbrace.
\end{eqnarray*}
Note that the inverse of Equation (\ref{alef}) is   
\begin{equation}
\label{winv}
\bm{y} =\bm{w}^{-1}_{\alpha}(\bm{m}) = \left (\frac{(1+\alpha m_1)^{1/\alpha}}{\sum_{j=1}^D (1+\alpha m_j)^{1/\alpha}},\ldots,\frac{(1+\alpha m_D)^{1/\alpha}}{\sum_{j=1}^D (1+\alpha m_j)^{1/\alpha}} \right ),
\end{equation}
for $\bm{m} \in \mathbb{Q}_{\alpha}^{D-1}$. As $\alpha \rightarrow 0$, Equation (\ref{alef}) converges to Equation (\ref{clr}) and Equation (\ref{winv}) becomes
\begin{equation}
\label{winv0}
\bm{y} =\bm{w}^{-1}_0(\bm{m}) = \left (\frac{e^ {m_1}}{\sum_{j=1}^D e^{m_j}},\ldots,\frac{e^ {m_D}}{\sum_{j=1}^D e^{m_j}} \right ).
\end{equation}
Finally, the $\alpha$--transformation is defined as
\begin{eqnarray} \label{alpha}
\bm{z}_{\alpha}(\bm{y})=\bm{H}\bm{w}_{\alpha}(\bm{y}).
\end{eqnarray}
The transformation in Equation (\ref{alpha}) is a one--to--one transformation which maps data inside the simplex onto a subset of $\mathbb{R}^{D-1}$ and vice versa for 
$\alpha \neq 0$. The corresponding sample space of Equation (\ref{alpha}) is 
\begin{eqnarray} \label{Ad}
\mathbb{A}_{\alpha}^{D-1}=\left\lbrace\bm{Hw}_{\alpha}(\bm{y}) \bigg | -\frac{1}{\alpha} \leq w_{i,\alpha} \leq \frac{D-1}{\alpha},\sum_{i=1}^Dw_{i, \alpha}=0 \right\rbrace.
\end{eqnarray}
The inverse transformation from $\mathbb{A}_{\alpha}^{D-1}$ to $\mathbb{S}^{D-1}$ is $\bm{z}_\alpha^{-1}(\bm{y}) = \bm{w}^{-1}_\alpha(\bm{H}^\top\bm{y})$ where $\bm{w}^{-1}(\cdot)$ is given in Equation (\ref{winv}). Note that vectors in $\mathbb{A}_{\alpha}^{D-1}$ are not subject to the sum--to--zero constraint and that $\lim_{\alpha \rightarrow 0}\mathbb{A}_{\alpha}^{D-1} \rightarrow \mathbb{R}^{D-1}$.

In effect, $\bm{y}_{\alpha}$ resembles a Box--Cox style mapping, and the resulting $\bm{y}_{\alpha}$ is an unconstrained vector in Euclidean space, suitable for standard multivariate statistical techniques. For convenience purposes we allow $\alpha$ to lie within $\left[-1,1\right]$. From Equations (\ref{stayalpha}) and (\ref{alef}), when $\alpha=1$, the simplex is linearly expanded as the values of the components are simply multiplied by a scalar and then centered. In this case, the transformation corresponds (up to scaling) to RDA. When $\alpha=-1$, the inverse of the values of the components are multiplied by a scalar and then centered. Thus, the transformation is aligned with RDA as well, but using the inverse of the compositional data. At the limiting case, as $\alpha \to 0$, the transformation converges to the ilr transformation used in LRA. Thus, the $\alpha$--transformation provides a continuum between RDA and LRA, allowing analysts to choose the most appropriate representation of compositional data based on empirical performance or theoretical considerations.

\subsection{The $\alpha$--regression}
In order to choose the optimal value of $\alpha$ \cite{tsagris2011} assumed that the $\alpha$--transformed data follow a multivariate normal distribution. However, the resulting space $\mathbb{A}_{\alpha}^{D-1}$ (\ref{Ad}) is a subset of $\mathbb{D}^{D-1}$ and \cite{tsagris2020} addressed this issue by proposing a folded multivariate normal distribution. The problem of zero values still remains, as the Jacobian of the $\alpha$--transformation contains the term $\sum_{ij}\log{y_{ij}}$. To overcome both problems, \cite{tsagris2015b} linked the compositional data to some predictors using the exponential link and employed the Kullback--Leibler divergence (KLD) to choose the optimal value of $\alpha$.

The $\alpha$--regression has the potential to improve the regression predictions with compositional data by adapting the $\alpha$--transformation to the dataset’s geometry. We assume that the conditional mean of the observed composition can be written as a non--linear function of some predictors
\begin{equation} \label{fit}
\mu_i = \begin{cases}
\displaystyle\frac{1}{1 + \sum_{j=1}^{D} e^{\bm{x}^\top \bm{\beta}_j}} & \text{for } i = 1\\[3ex]
\displaystyle\frac{e^{\bm{x}^\top \bm{\beta}_i}}{1 + \sum_{j=1}^{D} e^{\bm{x}^\top \bm{\beta}_j }} & \text{for } i = 2, \ldots, D
\end{cases}
\end{equation}
where 
\begin{eqnarray*} 
\pmb{\beta}_i=\left(\beta_{0i},\beta_{1i},...,\beta_{pi} \right)^\top, \ i=1,...,D-1 \ \ \text{and $p$ denotes the number of covariates}.
\end{eqnarray*}

\cite{tsagris2015b} used the log--likelihood of the multivariate normal distribution, but in this paper the regression is formulated as a non--linear least squares problem, where the minimizing function is the sum of squares of the errors (SSE)
\begin{eqnarray} \label{sse}
\text{SSE}\left(\bm{Y},\bm{X};\alpha, \bm{B}\right) = \sum_{j=1}^n\|\bm{y}_{j,\alpha}-\bm{\mu}_{j,\alpha}\|_2^2=\sum_{j=1}^n\left(\bm{y}_{j,\alpha}-\bm{\mu}_{j,\alpha}\right)^\top\left(\bm{y}_{j,\alpha}-\bm{\mu}_{j,\alpha}\right),
\end{eqnarray}
where $\bm{y}_{j,\alpha}$  and $\bm{\mu}_{j,\alpha}$ are the $\alpha$--transformations applied to the $j$--th observation and fitted compositional vectors, respectively, and $\|.\|_2$ denotes the $L_2$ norm. Application of the stay--in--the--simplex power transformation (\ref{stayalpha}) to the fitted vectors yields a simplified expression
\begin{eqnarray*}
\frac{\mu_i^{\alpha}}{\sum_{j=1}^D\mu_j^{\alpha}} = \frac{\left(\frac{e^{{\bf x}^\top\bm{\beta}_i}}{1+\sum_{j=1}^De^{{\bf x}^\top\bm{\beta}_j}}\right)^\alpha}{\frac{1+\sum_{k=1}^D\left(e^{{\bf x}^\top\bm{\beta}_k}\right)^\alpha}{\left(1+\sum_{j=1}^De^{{\bf x}^\top\bm{\beta}_j}\right)^\alpha}}=\frac{\left(e^{{\bf x}^\top\bm{\beta}_i}\right)^\alpha}{1 + \sum_{j=1}^D\left(e^{{\bf x}^\top\bm{\beta}_i}\right)^\alpha}.
\end{eqnarray*}

\subsubsection{Limiting case of $\alpha \rightarrow 0$} \label{sec::limit}
\cite{tsagris2016} presented the proof that as $\alpha \rightarrow 0$, the $\alpha$--transformation (\ref{alpha}) converges to the ilr transformation (\ref{ilr}). Following similar calculations one can show that 
\begin{eqnarray*}
\lim_{\alpha \rightarrow 0}{\frac{1}{\alpha} \left(D \frac{\mu_i^\alpha}{\sum_{j=1}^D\mu_j^\alpha} - 1\right) } \rightarrow \bm{x}\beta_i - \frac{\sum_{j=1}^D\bm{x}\beta_j}{D},
\end{eqnarray*}
which corresponds to the regression after the clr transformation (the ilr transformation (\ref{ilr}) without the right multiplication by the Helmert matrix). This implies that there are $D$ vectors of $\bm{\beta}$ regression coefficients. Due to the zero--sum of the set of regression coefficients, if we subtract the first coefficient from the rest of the $\bm{\beta}$ coefficients we end up with the regression coefficients of the additive log--ratio (alr) regression
\begin{eqnarray*}
\log\left(\frac{y_i}{y_1}\right)=\bm{x}^\top\pmb{\beta}_i, \ \ i=2,\ldots, D.
\end{eqnarray*}

\subsection{Choice of $\alpha$}
In the regression setting the optimal value of $\alpha$ is data--driven, and there are two ways to estimate its value. The first is to minimize the KLD between the observed and fitted compositions, $\text{KLD}\left(\bm{y},\bm{\mu} \right)=\sum_{j=1}^n\sum_{i=1}^Dy_{ij}\log{\left(y_{ij}/\mu_{ij}\right)}$. This results in a double minimization problem. For a given value of $\alpha$, one must minimize the SSE (\ref{sse}) in order to obtain the regression coefficients and then minimize the KLD with respect to $\alpha$ to obtain the optimal value of $\alpha$. With the choice of the KLD, the value of $\alpha$ is independent of the SSE, since the SSE is not comparable across the different values of $\alpha$. The second option is to examine $\alpha$ as a hyper--parameter whose value is chosen by minimizing the KLD via CV, e.g. 10--fold CV. 

\subsection{Consistency and asymptotic normality}

To establish the large--sample properties of the $\alpha$--regression estimator $\thetahat$, we impose a set of regularity conditions, organized into two groups. The first four assumptions (Assumptions~\ref{ass:dgp}--\ref{ass:continuity}) are sufficient for consistency, while the full set of seven assumptions (Assumptions~\ref{ass:dgp}--\ref{ass:nonsingular}) is required for asymptotic normality.

Recall that the estimator is defined as
\[
  \thetahat = \mathop{\arg\min}_{\theta \in \Theta}\; Q_n(\theta), \qquad
  Q_n(\theta) = \frac{1}{n}\sum_{j=1}^n \norm{\bm{y}_{j,\alpha} - \bm{\mu}_{j,\alpha}(\theta)}^2,
\]
and the population criterion is $Q(\theta) = \E\!\left[\norm{\bm{y}_{j,\alpha} - \bm{\mu}_{j,\alpha}(\theta)}^2\right]$.
The parameter vector $\theta = \operatorname{vec}(\bm{B}) \in \R^{d(p+1)}$ stacks the $d = D-1$ vectors of regression coefficients $\bm{\beta}_1, \ldots, \bm{\beta}_d$, each of length $p+1$ (including an intercept).

\begin{assumption}[Data generating process and correct specification]\label{ass:dgp}
The pairs $\{(\bm{y}_j, \bm{x}_j)\}_{j=1}^n$ are independent and identically distributed (i.i.d.) draws from a common distribution, and the conditional mean of the $\alpha$--transformed response is correctly specified:
\[
  \E\!\left(\bm{y}_{j,\alpha} \mid \bm{x}_j\right) = \bm{\mu}_{j,\alpha}(\thetao).
\]
\end{assumption}

\begin{remark}[Role of Assumption~\ref{ass:dgp}: identification of the population minimum]\label{rem:dgp_consistency}
Correct specification has two consequences. First, it ensures that the population criterion $Q(\theta)$ is uniquely minimized at $\thetao$, the true parameter. To see this, expand $Q(\theta)$ via the bias--variance decomposition:
\begin{align*}
  Q(\theta)
  &= \E\!\left[\norm{\bm{y}_{j,\alpha} - \bm{\mu}_{j,\alpha}(\theta)}^2\right] \\
  &= \E\!\left[\norm{(\bm{y}_{j,\alpha} - \bm{\mu}_{j,\alpha}(\thetao))
      + (\bm{\mu}_{j,\alpha}(\thetao) - \bm{\mu}_{j,\alpha}(\theta))}^2\right] \\
  &= \underbrace{\E\!\left[\norm{\bm{y}_{j,\alpha} - \bm{\mu}_{j,\alpha}(\thetao)}^2\right]}_{= Q(\thetao)}
     + \underbrace{\E\!\left[\norm{\bm{\mu}_{j,\alpha}(\thetao) - \bm{\mu}_{j,\alpha}(\theta)}^2\right]}_{\geq\, 0}
     + 2\underbrace{\E\!\left[(\bm{y}_{j,\alpha} - \bm{\mu}_{j,\alpha}(\thetao))^\top
       (\bm{\mu}_{j,\alpha}(\thetao) - \bm{\mu}_{j,\alpha}(\theta))\right]}_{=\, 0},
\end{align*}
where the cross--term vanishes by the tower property of conditional expectation and correct specification:
\[
  \E\!\left[(\bm{y}_{j,\alpha} - \bm{\mu}_{j,\alpha}(\thetao))^\top
    (\bm{\mu}_{j,\alpha}(\thetao) - \bm{\mu}_{j,\alpha}(\theta))\right]
  = \E\!\left[\E\!\left[\bm{y}_{j,\alpha} - \bm{\mu}_{j,\alpha}(\thetao)\mid \bm{x}_j\right]^\top
    (\bm{\mu}_{j,\alpha}(\thetao) - \bm{\mu}_{j,\alpha}(\theta))\right] = 0.
\]
Hence $Q(\theta) = Q(\thetao) + \E\!\left[\norm{\bm{\mu}_{j,\alpha}(\thetao)-\bm{\mu}_{j,\alpha}(\theta)}^2\right] \geq Q(\thetao)$ for all $\theta$. Under Assumption~\ref{ass:identification}, equality holds only at $\theta = \thetao$. Second, for the asymptotic normality proof, correct specification implies $\E[\bm{r}_j(\thetao)\mid\bm{x}_j]=\bm{0}$, which causes the second--order Hessian term to vanish (see Remark~\ref{rem:spec_norm} below). If the mean is misspecified, $\thetahat$ converges to the pseudo--true value $\theta^* = \arg\min_\theta Q(\theta)$, which in general differs from $\thetao$, inducing asymptotic bias.
\end{remark}

\begin{assumption}[Compact parameter space]\label{ass:compact}
The parameter space $\Theta \subset \R^{d(p+1)}$ is compact, and the true parameter satisfies $\thetao \in \operatorname{int}(\Theta)$.
\end{assumption}

\begin{remark}[Role of Assumption~\ref{ass:compact}: existence of estimator and ULLN]
Compactness of $\Theta$ serves two purposes. First, it guarantees that the minimizer $\thetahat$ exists for every $n$: a continuous function on a compact set attains its infimum (Weierstrass extreme value theorem; see, e.g., \citealt[Theorem~2.43]{rudin1976principles}). Second, compactness is the key condition under which the uniform law of large numbers (ULLN) converts pointwise convergence of $Q_n(\theta)$ into uniform convergence over $\Theta$ \citep[Lemma~2.4]{newey1994large}. Requiring $\thetao\in\operatorname{int}(\Theta)$ is indispensable for asymptotic normality: it ensures that the first--order condition $\nabla Q_n(\thetahat)=\bm{0}$ holds as an exact equality (rather than an inequality at a boundary), which is the starting point for the Taylor expansion in the proof of Theorem~\ref{theorem:normality}.
\end{remark}

\begin{assumption}[Identification]\label{ass:identification}
The population objective function $Q(\theta) = \E[\norm{\bm{y}_{j,\alpha} - \bm{\mu}_{j,\alpha}(\theta)}^2]$ is uniquely minimized at $\thetao$: for every $\epsilon > 0$ there exists $\delta > 0$ such that
\[
  \inf_{\theta:\,\norm{\theta-\thetao}\geq\epsilon} Q(\theta)
  \;\geq\; Q(\thetao) + \delta.
\]
\end{assumption}

\begin{remark}[Role of Assumption~\ref{ass:identification}: well--separated minimum]
The identification condition, sometimes called \emph{well--separation} \citep[Definition~5.1]{vaart1998asymptotic}, is what converts uniform convergence of $Q_n$ to $Q$ into convergence of $\thetahat$ to $\thetao$. Without a strict, separated gap, the minimizer of $Q_n$ could hover near a flat region of $Q$ without being driven to $\thetao$. As noted in Remark~\ref{rem:dgp_consistency}, correct specification alone guarantees $Q(\theta)\geq Q(\thetao)$, but not strict separation; Assumption~\ref{ass:identification} provides the additional strict inequality. In practice, well--separation is linked to the non--collinearity of the covariates and the injectivity of the map $\theta\mapsto\bm{\mu}_{j,\alpha}(\theta)$: if distinct parameters produce identical mean functions, the objective $Q$ would be flat and identification would fail.
\end{remark}

\begin{assumption}[Continuity and moment conditions]\label{ass:continuity}
The mean function $\bm{\mu}_{j,\alpha}(\theta)$ is continuous in $\theta$ for every $\theta \in \Theta$. Furthermore, the following second--moment bounds hold:
\[
  \E\!\left[\norm{\bm{y}_{j,\alpha}}^2\right] < \infty
  \qquad \text{and} \qquad
  \E\!\left[\sup_{\theta \in \Theta} \norm{\bm{\mu}_{j,\alpha}(\theta)}^2\right] < \infty.
\]
\end{assumption}

\begin{remark}[Role of Assumption~\ref{ass:continuity}: continuity and uniform integrability]
Continuity of $\bm{\mu}_{j,\alpha}(\theta)$ in $\theta$, combined with compactness of $\Theta$ (Assumption~\ref{ass:compact}), implies that $\theta \mapsto \norm{\bm{y}_{j,\alpha}-\bm{\mu}_{j,\alpha}(\theta)}^2$ is a continuous function of $\theta$ for each fixed sample path. This is a standard requirement for the ULLN \citep[Theorem~2.1]{newey1994large}: the class of functions $\{\theta \mapsto \norm{\bm{y}_{j,\alpha}-\bm{\mu}_{j,\alpha}(\theta)}^2 : \theta \in \Theta\}$ must be dominated by an integrable envelope. The condition $\E[\sup_\Theta\norm{\bm{\mu}_{j,\alpha}(\theta)}^2]<\infty$ provides this envelope, using the inequality
\[
  \sup_{\theta\in\Theta}\norm{\bm{y}_{j,\alpha}-\bm{\mu}_{j,\alpha}(\theta)}^2
  \;\leq\; 2\norm{\bm{y}_{j,\alpha}}^2 + 2\sup_{\theta\in\Theta}\norm{\bm{\mu}_{j,\alpha}(\theta)}^2,
\]
whose expectation is finite by the two moment bounds. For the finiteness of $\E[\norm{\bm{y}_{j,\alpha}}^2]$: although $\bm{y}_j\in\mathbb{S}^{D-1}$ is bounded, its $\alpha$--transformation $\bm{y}_{j,\alpha}$ need not be. Specifically, for $\alpha < 0$, the transformation involves $y_{ij}^{-|\alpha|}$, which diverges as a component $y_{ij}\to 0$. Thus $\E[\norm{\bm{y}_{j,\alpha}}^2]<\infty$ requires the distribution of $\bm{y}_j$ to place sufficiently little mass near the boundary of the simplex, and constitutes a genuine restriction on the data--generating process.
\end{remark}

\begin{theorem}[Consistency] \label{theorem:consistency}
Under Assumptions~\ref{ass:dgp}--\ref{ass:continuity},
\[
  \thetahat \convp \thetao \qquad \text{as } n \to \infty.
\]
\end{theorem}

The proof is given in Appendix~\ref{app:proof_consistency}.

\medskip
The next three assumptions are required for asymptotic normality.

\begin{assumption}[Smoothness]\label{ass:smoothness}
The mean function $\bm{\mu}_{j,\alpha}(\theta)$ is twice continuously differentiable in $\theta$ on $\operatorname{int}(\Theta)$. Define the \emph{negative Jacobian} (score matrix) as
\[
  g_j(\theta) = -\frac{\partial \bm{\mu}_{j,\alpha}(\theta)}{\partial \theta^\top}
  \in \R^{(D-1)\times d(p+1)},
\]
so that the gradient of the individual criterion is $\nabla_\theta f_j(\theta) = 2\,g_j(\theta)^\top \bm{r}_j(\theta)$, where $\bm{r}_j(\theta) = \bm{y}_{j,\alpha} - \bm{\mu}_{j,\alpha}(\theta)$ is the residual vector.
\end{assumption}

\begin{remark}[Verification of Assumption~\ref{ass:smoothness} for the $\alpha$--regression]
The fitted mean in~\eqref{fit} is a composition of exponential and rational functions of $\theta$, and is therefore infinitely differentiable in $\theta$ for all finite $\theta$. The explicit Jacobian, computed from the softmax structure of~\eqref{fit}, is supplied to the Levenberg--Marquardt algorithm to speed up computation (see Section~\ref{sec::app}). Thus Assumption~\ref{ass:smoothness} is automatically satisfied in the present model.
\end{remark}

\begin{assumption}[Strengthened moment conditions]\label{ass:moments}
The following uniform integrability conditions hold:
\[
  \E\!\left[\sup_{\theta \in \Theta} \norm{g_j(\theta)}^2\right] < \infty
  \qquad\text{and}\qquad
  \E\!\left[\sup_{\theta \in \Theta} \norm{\nabla^2_\theta f_j(\theta)}\right] < \infty,
\]
where $\norm{\cdot}$ denotes the Frobenius norm for matrices.
\end{assumption}

\begin{remark}[Role of Assumption~\ref{ass:moments}: ULLN for gradient and Hessian]
The condition on $g_j$ allows the USLLN to be applied to the gradient $\nabla Q_n(\theta) = \frac{2}{n}\sum_j g_j(\theta)^\top \bm{r}_j(\theta)$, establishing its uniform convergence to $\nabla Q(\theta)$. The condition on the Hessian $\nabla^2_\theta f_j(\theta)$ provides the integrable envelope needed to apply the mean value theorem uniformly in Step~2 of the proof of Theorem~\ref{theorem:normality}, specifically to control the remainder term in the second--order Taylor expansion. In both cases the conditions are uniform--in--$\theta$ analogues of the standard second--moment conditions used in the i.i.d.\ central limit theorem \citep[Lemma~3.1]{newey1994large}.
\end{remark}

\begin{assumption}[Non--singularity of the Gram matrix]\label{ass:nonsingular}
The \emph{Gram matrix}
\[
  \GMat = \E\!\left[g_j(\thetao)^\top g_j(\thetao)\right] \in \R^{d(p+1)\times d(p+1)}
\]
is positive definite (and hence invertible).
\end{assumption}

\begin{remark}[Role of Assumption~\ref{ass:nonsingular}: invertibility of the asymptotic Hessian]
Under Assumption~\ref{ass:dgp}, the Hessian of the population criterion evaluated at $\thetao$ satisfies
\[
  \nabla^2_\theta Q(\thetao)
  = 2\E\!\left[g_j(\thetao)^\top g_j(\thetao)\right]
    - 2\E\!\left[\nabla_{{\theta}} g_j(\thetao)^\top \bm{r}_j(\thetao)\right]
  = 2\GMat,
\]
because $\E[\bm{r}_j(\thetao)\mid \bm{x}_j]=\bm{0}$ (correct specification) causes the second term to vanish. Thus $\GMat$ equals half the population Hessian, which must be invertible to solve the linearized first--order condition in the proof of Theorem~\ref{theorem:normality}. Positive definiteness also ensures that $\thetao$ is a strict local minimum of $Q$, consistent with Assumption~\ref{ass:identification}. A sufficient condition for $\GMat\succ 0$ is that the columns of $g_j(\thetao)$ are not linearly dependent with positive probability, which is guaranteed if the covariates $\bm{x}_j$ are not (asymptotically) collinear.
\end{remark}

\begin{theorem}[Asymptotic normality] \label{theorem:normality}
Under Assumptions~\ref{ass:dgp}--\ref{ass:nonsingular}, as $n \to \infty$:
\begin{equation} \label{sandwich}
  \sqrt{n}(\thetahat - \thetao) \convd N\!\left(\bm{0},\; \GMat^{-1}\OmegaMat\GMat^{-1}\right),
\end{equation}
where
\[
  \GMat = \E\!\left[g_j(\thetao)^\top g_j(\thetao)\right],
  \qquad
  \OmegaMat = \E\!\left[g_j(\thetao)^\top \bm{r}_j(\thetao)\,\bm{r}_j(\thetao)^\top g_j(\thetao)\right].
\]
The matrix $\GMat^{-1}\OmegaMat\GMat^{-1}$ is the \emph{sandwich covariance matrix}.
\end{theorem}

The proof is given in Appendix~\ref{app:proof_normality}.

\begin{remark}[Sandwich covariance and the role of correct specification]\label{rem:spec_norm}
The covariance matrix $\GMat^{-1}\OmegaMat\GMat^{-1}$ is the \emph{Eicker--Huber--White sandwich estimator} \citep{white1980heteroskedasticity}, valid under heteroskedastic errors. Under the additional assumption that the errors $\bm{r}_j(\thetao)$ are conditionally homoskedastic---that is, $\E[\bm{r}_j(\thetao)\bm{r}_j(\thetao)^\top\mid\bm{x}_j]=\sigma^2 I_{D-1}$---one can show $\OmegaMat = \sigma^2 \GMat$, so the sandwich reduces to $\sigma^2\GMat^{-1}$, recovering the classical NLLS covariance. More generally, correct specification forces the cross--term $\E[\nabla g_j(\thetao)^\top\bm{r}_j(\thetao)]=\bm{0}$, so the ``bread'' simplifies to $\GMat$; without it, additional second--derivative terms appear in the asymptotic covariance \citep[Chapter~8]{wooldridge2010econometric}.
\end{remark}

\begin{remark}[Consistent estimation of the sandwich covariance]
In practice, $\GMat$ and $\OmegaMat$ are replaced by their sample analogues:
\[
  \widehat{G} = \frac{1}{n}\sum_{j=1}^n g_j(\thetahat)^\top g_j(\thetahat),
  \qquad
  \widehat{\Omega} = \frac{1}{n}\sum_{j=1}^n
    g_j(\thetahat)^\top \hat{\bm{r}}_j\,\hat{\bm{r}}_j^\top g_j(\thetahat),
\]
where $\hat{\bm{r}}_j = \bm{y}_{j,\alpha} - \bm{\mu}_{j,\alpha}(\thetahat)$. By the USLLN and Theorem~\ref{theorem:consistency}, $\widehat{G}\convp\GMat$ and $\widehat{\Omega}\convp\OmegaMat$. As noted in Section~\ref{sec:perm}, simulation studies indicate that permutation--based tests outperform the Wald test in finite samples, suggesting that the sandwich approximation may be slow to reach its asymptotic regime for compositional data.
\end{remark}

\subsection{Hypothesis testing} \label{sec:perm}
To conduct hypothesis testing regarding the statistical significance of the regression coefficients we may rely on the Wald test using the sandwich covariance of (\ref{sandwich}). However, our simulation studies have shown that the asymptotic Wald test does not perform satisfactorily and hence we used a permutation--based test, that relies on the SSE of the model, a which is computationally heavy though. 

With a fixed value of $\alpha$, to test for zero slope regression coefficients (omnibus or global test), $\bm{B}_{-1}=\bm{0}$, i.e. excluding the first row that corresponds to the constant terms, the following steps are implemented:
\begin{enumerate}
\item Estimate the $SSE^{obs}$ using all predictor variables.
\item Permute the observations (row--wise) of the compositional responses to destroy the relationship with the predictor variables and estimate the $SSE^{r}$.
\item Repeat Step 2 $R$ times (we suggest $R = 499$ permutations as a means to keep the computational burden low) and compute the permutation--based $p$--value as
\begin{equation*}
p-value=\frac{\sum_{r=1}^R\left(SSE^{r} \leq SSE^{obs}\right)+1}{R+1}.
\end{equation*}
\end{enumerate}

With a fixed value of $\alpha$, to test for a specific set of regression coefficients (partial or local test), for instance $\bm{B}_k=\bm{0}$, the $k$--th row of $\bm{B}$, corresponding to the coefficients of the $k$--th predictor variable, the following steps are implemented: 
\begin{enumerate}
\item Fit two $\alpha$--regression models, one using all predictor variables (full model), and one excluding the predictor variables whose regression coefficients are to be tested (reduced model), respectively.
\item Compute $W^{obs}=SSE^0-SSE^1$, where $SSE^0$ and $SSE^1$ refer to the SSE of the reduced and the full model, respectively.
\item Permute the observations (row--wise) of the predictor variable(s) to be tested, fit the full model (with all variables present) and compute $W^r=SSE^{0,r}-SSE^1$.
\item Repeat Steps 2--3 $R$ times and compute the permutation--based $p$--value as
\begin{equation*}
p-value=\frac{\sum_{r=1}^R\left(W^r \leq W^{obs}\right)+1}{R+1}.
\end{equation*}
\end{enumerate}

\begin{remark} 
Under i.i.d.\ errors and $H_0$, the permutation--based $p$--value satisfies $P(p_{perm} \leq \alpha) = \alpha$ exactly for any $n$ and $R \to \infty$.
The permutation--based test is consistent against fixed alternatives as $n \to \infty$, since $W^{obs} \to \infty$ while the $W^r$ remain $O_p(1)$. No normality assumption is required, and hence the test is valid under any exchangeable error distribution. 
\end{remark}

\subsection{MEs}
To account for the difficult interpretation of the regression coefficients, the MEs, given below, may be used
\begin{eqnarray} \label{me}
\text{ME}_{ik}=\frac{\partial \mu_i}{\partial x_k} = 
\left\lbrace
\begin{array}{ll} 
-\mu_1 \sum_{j=1}^{d} \beta_{jk} \mu_{j+1}  &  \text{for} \ \ i = 1 \\
\mu_i \left( \beta_{i-1,k} - \sum_{j=1}^{d} \beta_{jk} \mu_{j+1} \right) & \text{for} \ \  i=2,\ldots,D
\end{array}
\right\rbrace,
\end{eqnarray}
where $\sum_{i=1}^D\frac{\partial \mu_i}{\partial x_k}=0$, because $\sum_{i=1}^D\mu_i=1$. The MEs show the expected change in the compositional responses at an infinitesimal change of each predictor variable. They sum to zero, because if all response components increase, at least one response component must decrease by the same amount so that the unity sum constraint is preserved. The average MEs (AME) across all observations are then computed as
\begin{equation*}
\text{AME}_k = \frac{1}{n} \sum_{j=1}^{n} \frac{\partial \mu_j}{\partial x_k}.
\end{equation*}
Standard errors can be computed via bootstrap or the delta method, accounting for estimation uncertainty in both $\widehat{\bm{\beta}}$, and $\widehat{\bm{\mu}}$.

\subsection{ICE plots}
Due to the lack of straightforward interpretation of the estimated regression coefficients of the $\alpha$--regression, following \cite{tsagris2023} we can employ ICE plots to visualize the (non--linear) effect of each predictor variable on the fitted compositions. These can be assessed in parallel with the plots of the MEs, for each predictor variable separately. 

\subsection{Computational enhancement of the $\alpha$--regression}
For a given value of $\alpha$, the matrix of the regression coefficients $\bm{B}=\left(\bm{\beta}_1,\ldots,\bm{\beta}_d\right)$ is estimated using a modification of the Levenberg--Marquardt algorithm\footnote{This algorithm interpolates between the Gauss-–Newton algorithm and the method of gradient descent.}. The \textit{R} package \textsf{minpack.lm} \citep{minpacklm2023} is employed to this end. The Newton--Raphson algorithm was implemented but exhibited slower convergence. To enhance the speed of the function \texttt{nls.lm()} in the \textit{R} package \textsf{minpack.lm} we passed the function to compute the Jacobian matrix as an argument\footnote{This led to substantial improvements in the computational cost because instead of computing the Jacobian matrix numerically it computes it exactly and faster. The function became three times faster using the real data from the examples. The speed--up factor increases with increasing sample size, going up to 5 with tens of thousands of observations.}.  

We also tested the computational cost of the $\alpha$--regression with large--scale datasets, at the order of millions. We tested the performance of a meta--analytic approach where the dataset is partitioned into blocks, multiple subsets. This technique resulted in estimates whose values were in close agreement with the ones obtained from applying the $\alpha$--regression to the full dataset. However, the computational improvement was small, yielding a reduction of only 10\%--20\%, with use of parallel programming. 

\subsection{Extensions of the $\alpha$--regression}
Having established the standard $\alpha$--regression model and its inferential properties, we now extend the framework to spatially dependent compositional data. In the following sections, we define four spatial extensions of the $\alpha$--regression. The models presented cover the Euclidean predictors case, and the inclusion of compositional predictors is straightforward and hence not covered.  

\subsubsection{Robust $\alpha$--regression}
The $\alpha$--regression is based on minimization of the $L_2$ norm (\ref{sse}). In a similar fashion one may choose to minimize the sum of the absolute deviations, yielding the $\alpha$--minimum absolute deviations ($\alpha$--MAD) regression
\begin{eqnarray*}
\text{MAD}\left(\bm{Y},\bm{X};\alpha, \bm{B}\right) = \sum_{j=1}^n\sum_{i=1}^d|y_{ij,\alpha}-\mu_{ij,\alpha}|.
\end{eqnarray*}

Following \cite{tsagris2025}, the $\alpha$--MAD regression was formulated as a univariate regression problem by using the vectorization operation for the responses and by constructing the design matrix in a suitable manner. To make the estimation efficient, the command \texttt{nlrq()} from the \textit{R} package \textsf{quantreg} \citep{quantreg2025} was utilized. This approach exhibits dependence on initialization and does not ensure convergence. Alternative models include alternative loss functions. Instead of the $L_2$ norm, one may use the $L_1$ norm $\sum_{i=1}^n\|\bm{y}_{i,\alpha}-\bm{\mu}_{i,\alpha}\|_1 $ leading to the $\alpha$--spatial median regression\footnote{One option for this regression is to use \textit{R}'s built--in optimizers or iteratively reweighted least squares. The second option is computationally expensive as well, given the complexity of the derivatives involved.}. Other options include Tukey's biweight loss function \citep{tukey1960}, Hampel's loss function \citep{hampel1974} or Barron's general loss function \citep{barron2019}. The aforementioned robustified regression models, among others, are also included by making use of the \textit{R} package \textsf{gslnls} \citep{gslnls2025}. A practical limitation of these approaches is the increased computational cost. 

\subsubsection{Continuous and compositional predictors}
For convenience purposes, and without loss of generality, we will consider the case of a single composition, denoted by $\bm{Z}$. The composition $\bm{Z}$ is first transformed using the $\alpha$--transformation (\ref{alpha}), and is hence denoted by $\bm{Z}_{\alpha}$. Principal component analysis (PCA) computes the eigenvectors $\bm{V}_{\alpha}$ of $\bm{Z}_{\alpha}$ and then the projections onto this orthonormal basis are computed, $\bm{S}_{\alpha} = \bm{Z}_{\alpha}\bm{V}_{\alpha}$. 

The fitted values are given by
\begin{equation} \label{pcafit}
\mu_i^{\alpha'} = \begin{cases}
\displaystyle\frac{1}{1 + \sum_{j=1}^{D} e^{\bm{x}^\top \bm{\beta}_j+\bm{s}_{\alpha'}^\top \bm{\gamma}_j}} & \text{for } i = 1\\[3ex]
\displaystyle\frac{e^{\bm{x}^\top \bm{\beta}_i+\bm{s}_{\alpha'}^\top \bm{\gamma}_i}}{1 + \sum_{j=1}^{D} e^{\bm{x}^\top \bm{\beta}_j+\bm{s}_{\alpha'}^\top \bm{\gamma}_j }} & \text{for } i = 2, \ldots, D.
\end{cases}
\end{equation}

The notation $\alpha'$ highlights that the value of $\alpha$ in the compositional predictors need not be the same as the one used when computing the SSE (\ref{sse}), and to clarify the difference, the new SSE may be written as
\begin{eqnarray} \label{sse2}
\text{SSE}\left(\bm{Y},\bm{X};\alpha, \alpha', \bm{B}\right) = \sum_{j=1}^n\left(\bm{y}_{j,\alpha}-\bm{\mu}_{j,\alpha}^{\alpha'}\right)^\top\left(\bm{y}_{j,\alpha}-\bm{\mu}_{j,\alpha}^{\alpha'}\right).
\end{eqnarray}

Unlike the $\alpha$--regression, the extension contains two independent $\alpha$ values, whose values can be chosen via the KLD as mentioned earlier. For simplicity and to reduce computational cost, we propose to use the same value of $\alpha$ on both sides. The extension of the $\alpha$--regression to multiple compositional predictors is straightforward. 

\subsubsection{Temporal $\alpha$--regression}
The $\alpha$--regression can be extended to accommodate temporal dependence by incorporating lagged compositional responses as predictors. This follows naturally from the compositional predictors framework of the previous subsection, treating $\bm{Y}_{t-1}, \bm{Y}_{t-2}, \ldots, \bm{Y}_{t-q}$ as lagged compositional predictors alongside the exogenous predictors $\bm{x}_t$.

For a time series of compositions $\bm{Y}_t$, $t=1, \ldots, T$, let $q$ denote the autoregressive lag order. Each lagged composition $\bm{Y}_{t-\ell}$, $\ell = 1, \ldots, q$, is first $\alpha$--transformed and then projected onto its principal components following (\ref{pcafit}). Denoting the score vectors by $\bm{S}_{\alpha', t-\ell} = \bm{Z}_{\alpha', t-\ell} \bm{V}_{\alpha'}$, the fitted values at time $t$ are
\begin{equation} \label{ar1}
\mu_{it} =
\begin{cases}
\dfrac{1}{1 + \sum_{j=1}^{D} \exp\left(\bm{x}_t^\top \bm{\beta}_j 
  + \sum_{\ell=1}^q \bm{s}_{\alpha', t-\ell}^\top \bm{\gamma}_j^{(\ell)}\right)} 
& i = 1 \\[1em]
\dfrac{\exp\left(\bm{x}_t^\top \bm{\beta}_i 
  + \sum_{\ell=1}^q \bm{s}_{\alpha', t-\ell}^\top \bm{\gamma}_i^{(\ell)}\right)}
{1 + \sum_{j=1}^{D} \exp\left(\bm{x}_t^\top \bm{\beta}_j 
  + \sum_{\ell=1}^q \bm{s}_{\alpha', t-\ell}^\top \bm{\gamma}_j^{(\ell)}\right)} 
& i = 2, \ldots, D,
\end{cases}
\end{equation}
where $\bm{\Gamma}^{(\ell)} = \left(\bm{\gamma}_1^{(\ell)}, \ldots, \bm{\gamma}_d^{(\ell)}\right)$ 
is the matrix of regression coefficients for the $\ell$--th lag. The parameters $\bm{B} = (\bm{\beta}_1, \ldots, \bm{\beta}_d)$ and $\left\{\bm{\Gamma}^{(\ell)}\right\}_{\ell=1}^q$ are estimated by minimizing the SSE over $t = q+1, \ldots, T$
\begin{equation*}
\mathrm{SSE}\left(\bm{Y}, \bm{X};\, \alpha, \alpha', \bm{B}, 
\bm{\Gamma}^{(1)}, \ldots, \bm{\Gamma}^{(q)}\right) 
= \sum_{t=q+1}^{T} \left(\bm{y}_{t,\alpha} - \bm{\mu}_{t,\alpha}\right)^\top\left(\bm{y}_{t,\alpha} - \bm{\mu}_{t,\alpha}\right).
\end{equation*}
As noted in the previous subsection, for simplicity and to reduce computational cost, we propose to use the same value of $\alpha$ on both sides.

\subsubsection{Natural splines}
The $\alpha$--regression assumes a generalised--linear relationship between the covariates and the $\alpha$--transformed compositional response. This assumption can be relaxed by replacing the term $\bm{x}^\top\bm{\beta}_i$ with an additive expansion of natural cubic splines, yielding a more flexible regression model. For each covariate $x_k$, $k = 1, \ldots, p$, let
\begin{equation*}
f_{ki}(x_k) = \boldsymbol{\phi}_k(x_k)^\top \boldsymbol{\theta}_{ki},
\end{equation*}
where $\boldsymbol{\phi}_k(x_k) = \bigl(\phi_{k1}(x_k), \ldots,
\phi_{kM_k}(x_k)\bigr)^\top$ is the natural cubic spline basis vector for
covariate $x_k$, with $M_k$ basis functions determined by the placement of
$K_k$ interior knots, giving $M_k = K_k + 2$ after imposing the natural
(linearity beyond boundary knots) constraints, and
$\boldsymbol{\theta}_{ki} \in \mathbb{R}^{M_k}$ is the corresponding vector of spline coefficients for component $i$. The spline bases are pre--computed and stacked into an augmented design matrix
\begin{equation*}
\tilde{\bm{X}} = \bigl[\bm{1},\; \Phi_1,\; \Phi_2,\; \ldots,\; \Phi_p\bigr]
\in \mathbb{R}^{n \times \tilde{p}},
\end{equation*}
where $[\Phi_k]_{im} = \phi_{km}(x_{ik})$ and
$\tilde{p} = 1 + \sum_{k=1}^p M_k$ is the total number of parameters.
Each basis matrix $\Phi_k$ is column--centred to avoid collinearity with the
intercept. 
 
The interior knots for each predictor $\bm{x}_k$ are placed at equally spaced
quantiles of the observed values of $\bm{x}_k$. The number of knots $K_k$ is treated as a hyper--parameter and chosen via the same cross--validation protocol used to select $\alpha$, with the KLD as the performance metric. In practice,
$K_k \in \{1, 2, 3, 4, 5\}$ is sufficient for most applications, which may be selected using the permutation--based test described above. Natural cubic splines are constructed in \texttt{R}'s built--in package \textsf{splines} via the function \texttt{ns()}.

\section{The $\alpha$--spatial regression models} \label{sec::spat}
In the following sections, we define four spatial extensions of the $\alpha$--regression. The models presented cover the Euclidean predictors case, and the inclusion of compositional predictors is straightforward and hence not covered.  

\subsection{The $\alpha$--SLX model}
The $\alpha$--SLX model extends the standard $\alpha$--regression by incorporating spatial spillover effects through the covariates. The fitted compositional values are given by:
\begin{equation} \label{aslxfit}
\mu_i = \begin{cases}
\displaystyle\frac{1}{1 + \sum_{j=1}^{D} e^{\bm{x}^\top \bm{\beta}_j + (\bm{W}\bm{x})^\top \bm{\gamma}_j}} & \text{for } i = 1\\[3ex]
\displaystyle\frac{e^{\bm{x}^\top \bm{\beta}_i + (\bm{W}\bm{x})^\top \bm{\gamma}_i}}{1 + \sum_{j=1}^{D} e^{\bm{x}^\top \bm{\beta}_j + (\bm{W}\bm{x})^\top \bm{\gamma}_j}} & \text{for } i = 2, \ldots, D.
\end{cases}
\end{equation}
The matrices of regression coefficients $\bm{B}=\left(\bm{\beta}_1,\ldots,\bm{\beta}_d\right)$ and $\bm{\Gamma}=\left(\bm{\gamma}_1,\ldots,\bm{\gamma}_d\right)$ are estimated in the same way as in the $\alpha$--regression, and $\bm{W}$ is the contiguity matrix explained below.

\subsubsection{The contiguity matrix}
Some researchers tend to compute the Euclidean distance between two pairs of latitude and longitude, $(\nu_i,v_i)$ and $(\nu_j,v_j)$, $d_{ij}=\sqrt{\left(\nu_i-\nu_j\right)^2 + \left(v_i-v_j\right)^2}$. There is a flaw with this approach which is highlighted by \citet[pg.~13]{mardia2000}. Consider, for instance, the case of two coordinates whose latitude (or longitude) values are $359^{\circ}$ and $1^{\circ}$. Using the previous simplistic approach yields a distance between the two values $359^{\circ}-1^{\circ}=358^{\circ}$, but the actual distance between them is only $2^{\circ}$. To account for this, the pair of coordinates must first be transformed into their Euclidean coordinates prior to computing the Euclidean distance.  

The locations (latitude and longitude) between a pair of observations, $(\nu_i, v_i)$ and $(\nu_j, v_j)$ are first mapped from their polar to their Cartesian coordinates (after transforming the degrees into radians) 
$$\bm{c}_i =\left( \cos(\nu_i), sin(\nu_i) \cos(v_i), \sin(\nu_i) \sin(v_i) \right) \ \text{and} \ \bm{c}_j =\left( \cos(\nu_j), sin(\nu_j) \cos(v_j), \sin(\nu_j) \sin(v_j) \right).$$
The Euclidean distance between $\bm{c}_i$ and $\bm{c}_j$ is 
\begin{eqnarray*}
d(\bm{c}_i,\bm{c}_j) = d_{ij}^2= \left\|\bm{c}_i-\bm{c}_j\right\|^2= \left\| \bm{c}_i\right\|^2 + \left\| \bm{c}_j\right\|^2 - 2 \bm{c}_i^\top \bm{c}_j = 2\left(1- \bm{c}_i^\top \bm{c}_j\right). 
\end{eqnarray*}
For the $i$-th location, compute the region with the $k$ nearest neighbors $\mathcal{C}_{ik}$ and zero the rest, that is 
\begin{equation}
\tilde{w}_{ij}= \begin{cases}
1/d_{ij}^2 & \text{if} \ j \in \mathcal{C}_{ik} \\
\tilde{w}_{ij}=0 & \text{else}.  
\end{cases}
\end{equation}
The $(i,j)$ elements of the contiguity matrix $\bm{W}$ are then defined as $w_{ij}=\frac{\tilde{w}_{ij}}{\sum_{j=1}^n\tilde{w}_{ij}}$. 

\subsubsection{Choosing $\alpha$ and $k$}
The choice of the optimal values of $\alpha$ and of $k$ is again data--driven and can be performed via CV, but this time the spatial 10--fold CV protocol is employed, where the metric of performance is again the KLD.  

The \textit{R} package \textsf{blockCV} \citep{blockcv2019} implements spatial CV techniques designed to address the spatial autocorrelation inherent in geographical data. Unlike the traditional 10--fold CV, which can lead to overly optimistic model performance estimates when data points are spatially clustered \citep{roberts2017}, the spatial version partitions data into spatially separated training and testing folds. This ensures that the testing data are spatially independent from the training data, providing more realistic assessments of model generalization to new geographic areas. 

\subsubsection{Spatial MEs}
The spatial MEs (SMEs) consist of three components, the direct, the indirect and total MEs. The following formulas are identical to the standard $\alpha$--regression MEs (\ref{me}), as they depend only on the $\bm{\beta}$ coefficients and do not involve spatial terms.

The direct SMEs measure the change in the compositional responses at an infinitesimal change in the predictor variables
\begin{equation*}
DSME_{ik} = \frac{\partial \mu_i}{\partial x_k} = \begin{cases}
-\mu_1 \sum_{j=1}^{d} \beta_{jk} \mu_{j+1} & \text{for } i = 1 \\
 \mu_i \left(\beta_{i-1,k} - \sum_{j=1}^{d} \beta_{jk} \mu_{j+1}\right) & \text{for } i=2,\ldots, D.
\end{cases}
\end{equation*}

The indirect (spillover) SMEs measure the impact of a change in the spatially--lagged covariate $(\bm{W}\bm{x})_k$ (i.e., the weighted average of neighboring values) on the local composition component $\mu_i$. They have the same functional form as the direct effects, with $\bm{\gamma}$ replacing $\bm{\beta}$. This structural symmetry reflects how spatial spillovers operate through the same multiplicative mechanism as direct effects.
\begin{equation*}
IDSME_{ik} = \frac{\partial \mu_i}{\partial (\bm{W}\bm{x})_k} = \begin{cases}
-\mu_1 \sum_{j=1}^{d} \gamma_{jk} \mu_{j+1} & \text{for } i = 1 \\
\mu_i \left(\gamma_{i-1,k} - \sum_{j=1}^{d} \gamma_{jk} \mu_{j+1}\right) & \text{for } i=2,\ldots,D.
\end{cases}
\end{equation*}

The total SMEs combine both direct and indirect SMEs representing the full impact of a simultaneous infinitesimal change in both local and neighboring covariate values.
\begin{equation*}
TSME_{ik}=\frac{\partial \mu_i}{\partial x_k} + \frac{\partial \mu_1}{\partial (\bm{W}\bm{x})_k} = \begin{cases}
-\mu_1 \sum_{j=1}^{d} (\beta_{jk} + \gamma_{jk}) \mu_{j+1} & \text{for } i = 1 \\
\mu_i \left[(\beta_{i-1,k} + \gamma_{i-1,k}) - \sum_{j=1}^{d} (\beta_{jk} + \gamma_{jk}) \mu_{j+1}\right] & \text{for } i = 2,\ldots, D.
\end{cases}
\end{equation*}

The sum of the SMEs across all components equals zero:
\begin{equation*}
\sum_{i=1}^{D} \frac{\partial \mu_i}{\partial x_k} = 0 \quad \text{and} \quad \sum_{i=1}^{D} \frac{\partial \mu_i}{\partial (\bm{W}\bm{x})_k} = 0
\end{equation*}
This ensures that the composition remains on the simplex after perturbations.

Direct and indirect effects share the same functional form, differing only in the coefficient vectors used ($\bm{\beta}$ vs $\bm{\gamma}$). The contiguity matrix $\bm{W}$ determines which neighbors contribute to spillover effects. We remind that row--standardization is used such that $\sum_j w_{ij} = 1$.

\subsubsection{Prediction of new values}
To predict the compositions for new observations $\bm{x}_{new}$, we must first construct the matrix $\bm{W}_{new}$ which contains the row normalized distances from the new locations to the existing ones, and then use the following formula
 \begin{equation} \label{slxpreds}
\widehat{\mu}_i = \begin{cases}
\displaystyle\frac{1}{1 + \sum_{j=1}^{D} e^{\bm{x}_{new}^\top \bm{\beta}_j + (\bm{W}_{new}\bm{x})^\top \bm{\gamma}_j}} & \text{for } i = 1\\[3ex]
\displaystyle\frac{e^{\bm{x}_{new}^\top \bm{\beta}_i + (\bm{W}_{new}\bm{x})^\top \bm{\gamma}_i}}{1 + \sum_{j=1}^{D} e^{\bm{x}_{new}^\top \bm{\beta}_j + (\bm{W}_{new}\bm{x})^\top \bm{\gamma}_j}} & \text{for } i = 2, \ldots, D.
\end{cases}
\end{equation}

\subsection{The $\alpha$--SAR model}
Inspired by the SAR for the Dirichlet regression \citep{nguyen2026modeling}, we define the following formulation. The fitted values are defined in a manner similar to (\ref{fit})
\begin{equation} \label{sarfit}
\mu_i = \begin{cases}
\displaystyle\frac{1}{1 + \sum_{j=1}^{D} e^{\bm{S}\left(\rho\right)^{-1}\bm{x}^\top \bm{\beta}_j}} & \text{for } i = 1\\[3ex]
\displaystyle\frac{e^{\bm{S}\left(\rho\right)^{-1}\bm{x}^\top \bm{\beta}_i}}{1 + \sum_{j=1}^{D} e^{\bm{S}\left(\rho\right)^{-1}\bm{x}^\top \bm{\beta}_j}} & \text{for } i = 2, \ldots, D,
\end{cases}
\end{equation}
where $\rho \in (-1,1)$ is the spatial autoregressive parameter measuring spillover strength, $\bm{S}\left(\rho\right)=\bm{I}_n-\rho\bm{W}$ is the spatial multiplier matrix. 

Similarly to the $\alpha$--regression, for a given value of $\alpha$ we minimize the SSE (\ref{sse}) in order to estimate the $\beta$s and the $\rho$ parameter. The choice of $\alpha$ and $k$ (number of nearest neighbours in $\bm{W}$) is again performed via the spatial 10--fold CV.

\subsubsection{SMEs}
The direct effects measure the impact of a change in location $i$'s covariate on location $i$'s own composition:
\begin{equation*}
DSME_{ilk}=\frac{\partial \mu_{i\ell}}{\partial x_{ik}} = 
\begin{cases}
-\mu_{i1}\sum_{j=1}^{d} \beta_{jk}\mu_{ij+1} \cdot [\bm{S}(\rho)^{-1}]_{ii} & \text{for } \ell = 1\\[2ex]
\mu_{i\ell}\left[\beta_{\ell-1,k} - \sum_{j=1}^{d} \beta_{jk}\mu_{ij+1}\right] \cdot [\bm{S}(\rho)^{-1}]_{ii} & \text{for } \ell = 2, \ldots, D.
\end{cases}
\end{equation*}

The indirect effect at location $i$ (summing spillovers from all neighbors) is
\begin{equation*}
IDSME_{iljk}=\frac{\partial \mu_{i\ell}}{\partial x_{jk}} = 
\begin{cases}
-\mu_{i1}\sum_{j=1}^{d} \beta_{jk}\mu_{ij+1} \cdot \sum_{j \neq i}[\bm{S}(\rho)^{-1}]_{ij} & \text{for } \ell = 1\\[2ex]
\mu_{i\ell}\left[\beta_{\ell-1,k} - \sum_{j=1}^{d} \beta_{jk}\mu_{ij+1}\right] \cdot \sum_{j \neq i}[\bm{S}(\rho)^{-1}]_{ij} & \text{for } \ell = 2, \ldots, D
\end{cases}
\end{equation*}

The total SMEs are the sum of the direct and indirect effects.

\subsubsection{Prediction of new values}
Denote the new $m$ covariate values by $\bm{X}^{new}$ located at new, unseen in the model, coordinates. We stack the new covariate values under the observed ones to create the augmented design matrix \citep{goulard2017predictions}
\begin{eqnarray*}
\bm{X}^{aug}=\left(
\begin{array}{cc}
 \bm{X} &
 \bm{X}^{new}     
\end{array}      
\right).
\end{eqnarray*}
Similarly define 
\begin{eqnarray*}
\bm{W}^{aug}=\left(
\begin{array}{cc}
 \bm{W} & \bm{W}^{new} \\
 \bm{W}^{new} & \bm{W}     
\end{array}      
\right)
\end{eqnarray*}
to be the augmented contiguity matrix, where $\bm{W}^{new}$ denotes the distances of the new locations from the observed ones. Note that $\bm{W}^{aug}$ is row standardised. $\bm{X}^{aug}$ contains $n+m$ rows, and $\bm{W}^{aug}$ is of dimensions $(n+m) \times (n+m)$, where $n$ is the sample size of the observed sample, upon which the estimates are derived, and $m$ denotes the size of the new observations (locations, sites).

The predicted values are given by 
\begin{equation*} 
\widehat{y}^{aug}_{i} = \begin{cases}
\displaystyle\frac{1}{1 + \sum_{\ell=1}^{D} e^{(\bm{I}_{n+m} - \rho \bm{W}^{aug})^{-1}\left(\bm{x}_j^{aug}\right)^\top \bm{\beta}_\ell}} & \text{for } i = 1\\[3ex]
\displaystyle\frac{e^{(\bm{I}_{n+m} - \rho \bm{W}^{aug})^{-1}\left(\bm{x}_j^{aug}\right)^\top \bm{\beta}_i}}{1 + \sum_{\ell=1}^{D} e^{(\bm{I}_{n+m} - \rho \bm{W}^{aug})^{-1}\left(\bm{x}_j^{aug}\right)^\top \bm{\beta}_\ell}} & \text{for } i = 2, \ldots, D.
\end{cases}
\end{equation*}
Stacking the predicted values, in a matrix format,
$
\widehat{\bm{Y}}^{aug}=\left(\begin{array}{c}
\widehat{\bm{Y}} \\
\widehat{\bm{Y}}^{new} 
\end{array}  \right),
$
we observe that the predictions for the new covariate values at the new locations are placed in the bottom $m$ rows of $\widehat{\bm{Y}}^{aug}$. 

\subsubsection{Computational challenges}
The main obstacle faced during estimation of the $\alpha$--SAR model is the inversion of the $n \times n$ matrix $\bm{S}\left(\rho\right)$, a task that becomes computationally heavier as the sample size increases. Second, prior to performing the Levenberg--Marquardt algorithm we perform a grid search of $\rho$ values, then estimate the parameters for a given value of $\rho$ and choose the $\rho$ that yields the minimum SSE. Each time, initial values for the $\beta$s are derived by the $\alpha$--regression. Then, we use this $\rho$ value and the resulting $\beta$s as starting values for the estimation of the model. 

\subsection{The GW$\alpha$R model}
The GW$\alpha$R model is a weighted $\alpha$--regression scheme, but the difference is that the regression is performed $n$ times, each time with different weights. The weighted $SSE$ that must be minimized is 
\begin{eqnarray} \label{wsse}
SSE\left(\bm{Y},\bm{X};\alpha, h, \bm{B}\right) = \sum_{j=1}^n\left(\bm{y}_{j,\alpha}-\bm{\mu}_{j,\alpha}\right)^\top \bm{W}_i\left(\bm{y}_{j,\alpha}-\bm{\mu}_{j,\alpha}\right),
\end{eqnarray}
where $\bm{W}_j=\text{diag}\left\{w_{j1},\ldots,w_{jn}\right\}$ is the weighting matrix corresponding to the weights allocated to the $j$--th observation. A common weighting function is the Gaussian kernel
\begin{eqnarray} \label{rbf}
w_{ij} = \exp \left( - \frac{d_{ij}^2}{2h^2} \right),
\end{eqnarray}
where $d_{ij}$ is the distance between location $i$ and $j$, and $h$ is the bandwidth parameter controlling the degree of spatial smoothing.

As $\alpha \rightarrow 0$, the GW$\alpha$R converges to the GWR under the alr transformation \citep{yoshida2021}.

\subsubsection{Choice of $\alpha$ and $h$}
Choosing the optimal value of $h$ in the classical GWR is typically achieved via the spatial 10--fold CV protocol, with the KLD acting as the metric of performance. The GW$\alpha$R model entails an extra hyper--parameter, $\alpha$. This time the CV protocol is computationally more intensive. To alleviate the cost, the range of possible values $\alpha$ to be examined may be reduced, retaining only distinct values such as $\alpha=0.1,0.25,0.5,0.75,1.0$. A heuristic approach to expedite the identification of the optimal $\alpha$ value involves performing the CV protocol using the $\alpha$--regression. However, empirical evidence suggests this strategy is inadvisable. Regarding the $h$ hyper--parameter, following \cite{gretton2012,schrab2023} the median distance heuristic is employed as the starting point. This way, one knows a region to search for the optimal value of $h$. 

\subsubsection{SMEs}
The formula for the SMEs of the GW$\alpha$R is nearly the same as that of the $\alpha$--regression (\ref{me}), but location--specific
\begin{equation} \label{wme}
\frac{\partial \mu\left(\nu_i, v_i\right)}{\partial x_k} = \begin{cases} 
-\mu_1\left(\nu_i, v_i\right) \sum_{j=1}^{d} \beta_{jk}\left(\nu_i, v_i\right) \mu_{j+1}\left(\nu_i, v_i\right) & \text{for } i = 1 \\
\mu_\ell\left(\nu_i, v_i\right) \left[ \beta_{i-1,k}\left(\nu_i, v_i\right) - \sum_{j=1}^{d} \beta_{jk}\left(\nu_i, v_i\right) \mu_{j+1}\left(\nu_i, v_i\right) \right] & \text{for } \ell=2,\ldots,D.
\end{cases}
\end{equation}
Similarly to the $\alpha$--regression, the $\sum_{\ell=1}^D\frac{\partial \mu_\ell\left(\nu_i, v_i\right)}{\partial x_k}=0$, but this time, this is true for every location. 

\subsubsection{Computational tricks to alleviate the computational burden}
The weighting function (\ref{rbf}) becomes $w_{ij} = \exp \left( - \frac{d_{ij}^2}{2h^2} \right) = \exp \left( \frac{\bm{c}_i^\top \bm{c}_j-1}{h^2} \right)$. The minimization of the $SSE$ takes place for specific values of $\alpha$ and $h$. When passing the arguments of the $SSE$ in the command \texttt{minpack.lm::nls.lm()}, the quantity $\alpha \bm{x}$ is pre--computed and passed as an argument. The function \texttt{minpack.lm::nls.lm()} requires a function that outputs the residuals. So, in order to perform weighted least squares we multiply the weights $\bm{w}_i$ by the residuals $\bm{r}_i$. Finally, for each observation $i$, we can compute the regression coefficients for different values of $h$. This is useful during the CV protocol.

\subsection{The $\alpha$--ESF model}
To overcome the computational challenges associated with the $\alpha$--SAR (and the GW$\alpha$R) model we propose the use of the $\alpha$--ESF model. In contrast to the $\alpha$--SLX model, the $\alpha$--ESF model requires a (symmetric and un--normalised) distance matrix $\bm{D}$, where the $D_{i,j}$ element denotes the spatial Euclidean distance between the sample sites $i$ and $j$. Then we create the kernelized matrix $\bm{C}$, using the exponential kernel where $\bm{C}_{i,j}=e^{-D_{i,j}/h}$, or the Gaussian kernel where $\bm{C}_{i,j}=e^{-D_{i,j}^2/h}$. The value of the parameter $h$ is equal to the maximum length of the minimum spanning tree connecting sample sites. The matrix $\bm{C}$ is doubly centered, $\bm{A}=\bm{MCM}$, where $\bm{M}=\bm{I}_n-\frac{1}{n}\bm{j}_n\bm{j}^\top$, with $\bm{j}_n$ indicating the $n$--dimensional vector of 1s. 

To compute the kernelized matrix we use the \textit{R} package \textsf{spmoran} \citep{spmoran2024}. The package allows for a user--specified distance matrix where we can pass the distance matrix computed using polar coordinates, the same as with the $\alpha$--SLX and GW$\alpha$R models, or allow the package to compute the Euclidean distances on the coordinates. This time we preferred the second choice because the package approximates the eigenvectors of the kernelised distance matrix computationally efficiently, even with large sample sizes \citep{murakami2019eigenvector}. 

The final step is to compute the eigenvectors $\bm{V}$ of $\bm{A}$ and use the $k$ eigenvectors corresponding to the $k$ largest positive eigenvalues as spatial predictors, and link them to the compositional responses using the same link as previously:
\begin{equation} \label{aesffit}
\mu_i = \begin{cases}
\displaystyle\frac{1}{1 + \sum_{j=1}^{D} e^{\bm{x}^\top \bm{\beta}_j +\bm{V}_k^\top \bm{\gamma}_j}} & \text{for } i = 1\\[3ex]
\displaystyle\frac{e^{\bm{x}^\top \bm{\beta}_i + \bm{V}^\top \bm{\gamma}_i}}{1 + \sum_{j=1}^{D} e^{\bm{x}^\top \bm{\beta}_j + \bm{V}_k^\top \bm{\gamma}_j}} & \text{for } i = 2, \ldots, D.
\end{cases}
\end{equation}

The regression coefficients of the eigenvectors are not interpreted in the traditional way and the SMEs and ICE plots are essential to visualize the effect of these coefficients. The same formulas as those provided for the $\alpha$--regression are applicable here.

\subsubsection{Choosing the appropriate number of eigenvectors}
A variable selection algorithm is required to select the number of eigenvectors. Due to the nature of the $\alpha$--regression we cannot use classical algorithms such as LASSO \citep{tibshirani1996} or likelihood--based stepwise algorithms. We choose to use the $\gamma$--OMP algorithm \citep{tsagris2022gamma} that was devised to work with any regression model. We start with the predictor variables alone and compute the KLD. Then, we apply the $\gamma$--OMP algorithm and select eigenvectors that reduce the KLD. Addition of not important eigenvectors may increase the KLD, at which point the algorithm terminates. 

The drawback of this criterion is the selection of too many eigenvectors (overfitting) and thus following the suggestion of \cite{tsagris2022gamma} we may substitute this criterion with a statistical test. Upon selecting a candidate for inclusion eigenvector, we can perform a permutation--based test (see Section \ref{sec:perm}) to determine whether this eigenvector is statistically significant at the 5\% significance level. The introduction of at least one eigenvector implies the presence of spatial dependence. To test for spatial dependence we rely on the permutation--based test described in Section \ref{sec:perm}.

\subsubsection{Effect of predictor variables and prediction of new values}
Regarding the SMEs and ICE plots we can rely on the same formulas as those provided for the $\alpha$--regression. Regarding prediction of the compositional responses for observations at new locations requires the Nystrom extension \citep{drineas2005nystrom} to estimate the spatial eigenvectors at the unobserved sites.

\subsection{Extensions}
The most natural extension is the spatial Durbin model (SDM) which combines the $\alpha$--SAR and $\alpha$--SLX models in one. In this case, the fitted values have a similar style formula:
\begin{equation*}
\mu_i = \begin{cases}
\displaystyle\frac{1}{1 + \sum_{j=1}^{D} e^{\bm{S}\left(\rho\right)^{-1}\left[\bm{x}^\top \bm{\beta}_j+(\bm{W}\bm{x})^\top \bm{\gamma}_j\right]}} & \text{for } i = 1\\[3ex]
\displaystyle\frac{e^{\bm{S}\left(\rho\right)^{-1}\left[\bm{x}^\top \bm{\beta}_i+(\bm{W}\bm{x})^\top \bm{\gamma}_i\right]}}{1 + \sum_{j=1}^{D} e^{\bm{S}\left(\rho\right)^{-1}\left[\bm{x}^\top \bm{\beta}_j+(\bm{W}\bm{x})^\top \bm{\gamma}_j\right]}} & \text{for } i = 2, \ldots, D.
\end{cases}
\end{equation*}
We can then define the $\alpha$--spatio--temporal model to combine the fitted values of the $\alpha$--SAR model (\ref{sarfit}) with the temporal $\alpha$--regression (\ref{ar1}) formulation of the fitted values and perhaps even include the $\alpha$--SLX model fitted values (\ref{aslxfit}), or combine the temporal $\alpha$--regression with the $\alpha$--SLX model. The drawback of these extensions is the computational challenges of the formulations. 

A perhaps better extension is the $\alpha$--ESF model where the spatio--temporal correlation is captured in the eigenvectors. The $\alpha$--ESF model further allows to approximate the spatially varying coefficients of the GW$\alpha$R model by incorporating the interaction between the spatial (or spatio--temporal) eigenvectors and the predictors. Finally, the addition of natural splines to the $\alpha$--ESF model can increase the non--linearity of the model.  

\section{Simulation studies} \label{sec:sims}
We examined the performance of the permutation--based hypothesis test for the regression coefficients of the $\alpha$--regression, in terms of type I error and power, for both the omnibus and the partial test, using 1,000 repetitions. The sample sizes considered were always $n=100,200,500$ and the number of compositional responses was set equal to 5, $D=5$. 

\subsection{Omnibus test}
For the omnibus test, we followed the following steps. In the case of independence, we generated random Dirichlet vectors for the compositional responses and $p$ random normal predictor variables, where $p=1,2,3,4$. To estimate the power, we again generated values for $p$ normal predictor variables and linked them to the response variable via the exponential link $\bm{\mu}=\left(1, e^{\bm{x}_1^\top \bm{\beta}_1}, e^{\bm{x}_D^\top \bm{\beta}_D}\right)^\top$. The compositional responses were then generated from a Dirichlet distribution, $\bm{y}_j \sim Dir(\bm{\mu}_j)$.

Table \ref{sizepow1} contains the results for the type I error and the power. The omnibus test is size correct regardless of the sample size and the number of predictor variables used. The estimated power is high and increases with increasing sample sizes, as expected, and with the number of predictor variables.

\begin{table}[ht]
\centering
\caption{Estimated type I error and power for the omnibus test for three sample sizes and different numbers of predictors ($p$) whose significance is to be tested.}
\label{sizepow1}
\begin{tabular}{r|rrrr|rrrr}
\toprule
       & \multicolumn{4}{c}{Type I error} & \multicolumn{4}{c}{Power} \\ \midrule 
Sample size & $p=1$ & $p=2$ & $p=3$ & $p=4$ & $p=1$ & $p=2$ & $p=3$ & $p=4$ \\  \midrule
$n=100$ & 0.062 & 0.040 & 0.048 & 0.048 & 0.774 & 0.926 & 0.868 & 0.996 \\ 
$n=200$ & 0.046 & 0.048 & 0.054 & 0.042 & 0.960 & 1.000 & 1.000 & 1.000\\ 
$n=500$ & 0.052 & 0.042 & 0.062 & 0.046 & 1.000 & 1.000 & 1.000 & 1.000\\ 
\bottomrule
\end{tabular}
\end{table}

\subsection{Partial test}
For the partial test, we followed the following steps. In the case of independence, we generated again $4$ normal predictor variables and linked them to the response variable via the exponential link $\bm{\mu}=\left(1, e^{\bm{x}_1^\top \bm{\beta}_1}, e^{\bm{x}_D^\top \bm{\beta}_D}\right)^\top$. The compositional responses were then generated from a Dirichlet distribution, $\bm{y}_j \sim Dir(\bm{\mu}_j)$. We then generated $p=1,2$ random normal variables and added them to the predictor variables. This way, the first 4 predictor variables are related to the response whereas the rest are not. To estimate the power of the test we simply followed the aforementioned technique, but with 5 or 6 predictor variables linked to the compositional responses and tested the significance of the 5th predictor variable (denoted by $p=1$ in Table \ref{sizepow2}) and of both the 5th and 6th predictor variables together (denoted by $p=2$ in Table \ref{sizepow2}).

Table \ref{sizepow2} contains the results for the type I error and the power. The partial test is size correct regardless of the sample size and the number of predictor variables used. The estimated power is high and increases with increasing sample sizes, as expected, and with the number of predictor variables. But this time, the estimated power is lower than that of the omnibus test, for the same sample sizes.

\begin{table}[ht]
\centering
\caption{Estimated type I error and power for the partial test for three sample sizes and the number of predictors ($p$) to be tested.}
\label{sizepow2}
\begin{tabular}{r|rr|rr}
\toprule
  & \multicolumn{2}{c}{Type I error} & \multicolumn{2}{c}{Power} \\ \midrule 
Sample size & $p=1$ & $p=2$ & $p=1$ & $p=2$ \\ \midrule
$n=100$ & 0.036 & 0.042 & 0.800 & 0.900 \\ 
$n=200$ & 0.041 & 0.060 & 0.838 & 0.926 \\ 
$n=500$ & 0.042 & 0.038 & 1.000 & 1.000 \\  
\bottomrule
\end{tabular}
\end{table}

\section{Application to real datasets} \label{sec::app}
Real--data applications show that the $\alpha$--regression can outperform the standard LRA--based regression, in terms of predictive performance, particularly when zeros are present, which can be further improved by taking into account the spatial dependencies. Four real datasets were considered, two that involved spatial dependence and two that contained compositional predictors. The experiments were performed on a Dell laptop with Intel Core i7-1355U (1.70 GHz), 16GB RAM, 512GB SSD, and Windows 11 Pro installed.

The first two datasets are considered, but ignoring the spatial dependence this time to serve two purposes: (a) visualization of the effect of the predictor variables on the fitted compositions and (b) comparison of the $\alpha$--regression to the $\alpha$--$k$--$NN$ regression of \citep{tsagris2025}. The third dataset was used to illustrate the performance of the $\alpha$--regression with compositional predictors and to compare it to the $\alpha$--simplicial constrained least squares ($\alpha$--SCLS) model of \cite{tsagris2025}. The fourth dataset concerns with temporal dependence. Information regarding these datasets is delineated below:
\begin{itemize}
\item \textbf{Agricultural economics dataset}: Data regarding crop productivity in the Greek NUTS II region of Thessaly during the 2017--2018 cropping year were supplied by the Greek Ministry of Agriculture. The data refer to a sample of farms and initially they consisted of 20 crops, but after grouping and aggregation they were aggregated into five crop categories\footnote{A larger version of this dataset was used in \cite{mattas2026}. Following the EU Regulation No1166/2008 that establishes a framework for European statistics at the level of agricultural holdings the aggregation took place across different output of crops.}. These crops are \textit{cereals}, \textit{cotton}, \textit{tree crops}, \textit{other annual crops and pasture} and \textit{grapes and wine}. For each of the 168 farms with unique coordinates, the cultivated area in each of these 5 grouped crops is known. The goal is to examine the relationship between the composition of the cultivated area and the following covariates: human influence index \textit{(HII}, direct human influence on ecosystems). Zero value represents no human influence and 64 represents maximum human influence possible. The other two covariates were the soil pH (\textit{CaCl$_2$}), and the topsoil organic carbon content (\textit{SOC}).

\item \textbf{Meuse river dataset}: This dataset gives locations and topsoil heavy metal concentrations, along with a number of soil and landscape variables at the observation locations, collected in a flood plain of the river Meuse, near the village of Stein (Netherlands). Heavy metal concentrations are from composite samples of a squared area of approximately 15m $\times$ 15m. There are measurements (all measured in mg kg$^{-1}$ (ppm)): \textit{topsoil cadmium concentration} (zero cadmium values in the original dataset have been shifted to 0.2 (half the lowest non--zero value)), \textit{topsoil copper concentration}, \textit{topsoil lead concentration} and \textit{topsoil zinc concentration}. We have selected 3 covariates to associate the components with, namely the \textit{relative elevation above local river bed} (in metres), the \textit{organic matter}, kg (100 kg)$^{-1}$ soil (percent) and the \textit{distance to river Meuse} (in metres), as obtained during the field survey. 

\item \textbf{Greek national elections dataset}: This dataset is used to examine the performance of the $\alpha$--regression with compositional predictors. The compositional response contains the percentages of votes at each NUTS--3 region\footnote{Greece consists of 63 such regions.} for each of the 8 mainstream political parties in Greece during the 2023 national elections. The political parties are \textit{ND}, \textit{SYRIZA}, \textit{PASOK}, \textit{KKE}, \textit{Spartans}, \textit{Greek Solution}, \textit{Victory} and \textit{Course Freedom}. The compositional predictor contains the percentage of labour allocated to each of the following 6 categories, \textit{agriculture}, \textit{industry}, \textit{construction}, \textit{trade \& tourism}, \textit{business \& finance}, and \textit{public services}. 

\item \textbf{Catalan elections}: The data set contains the votes in Catalan elections from year 1980 up to 2006 for 41 regions each year. The main parties consist of 6 candidates, while there are votes for other candidates, blank votes and null votes. In total there are 328 observations with 9 variables, that were scaled to sum to 1. The goal here is to assume an AR(1) model, where the lag refers to the one time period between two consecutive election years. 

\end{itemize}

Due to the presence of zero values the LRA approach, i.e. the family of the $\alpha$--regression models presented earlier with $\alpha=0$, is not applicable in the Agricultural economics and the Catalan elections datasets. 

\subsection{Inspection of the $\alpha$--regression, without assuming spatial dependence}
\begin{figure}[!ht]
\centering
\begin{tabular}{cc}
\includegraphics[scale = 0.3]{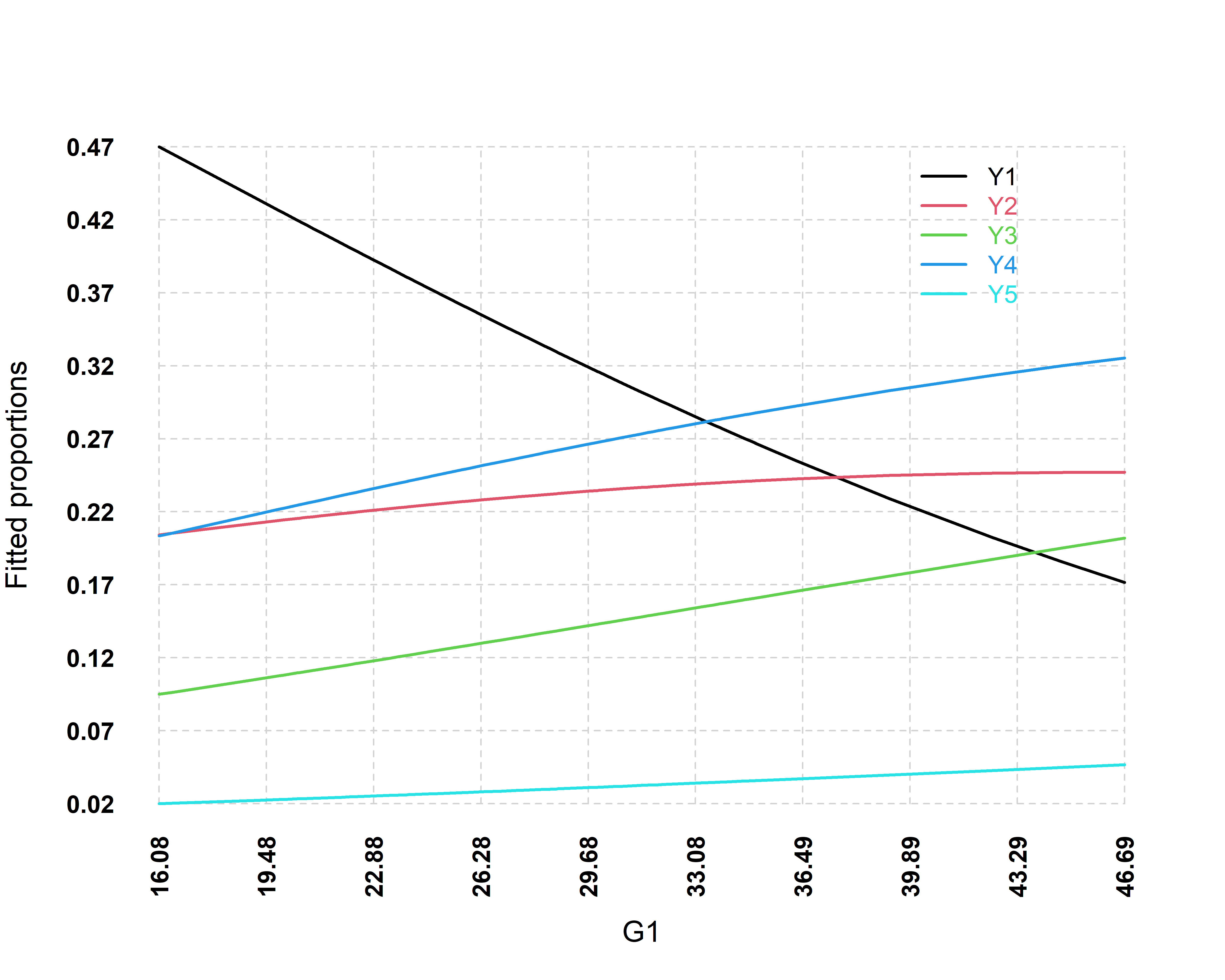}  &
\includegraphics[scale = 0.3]{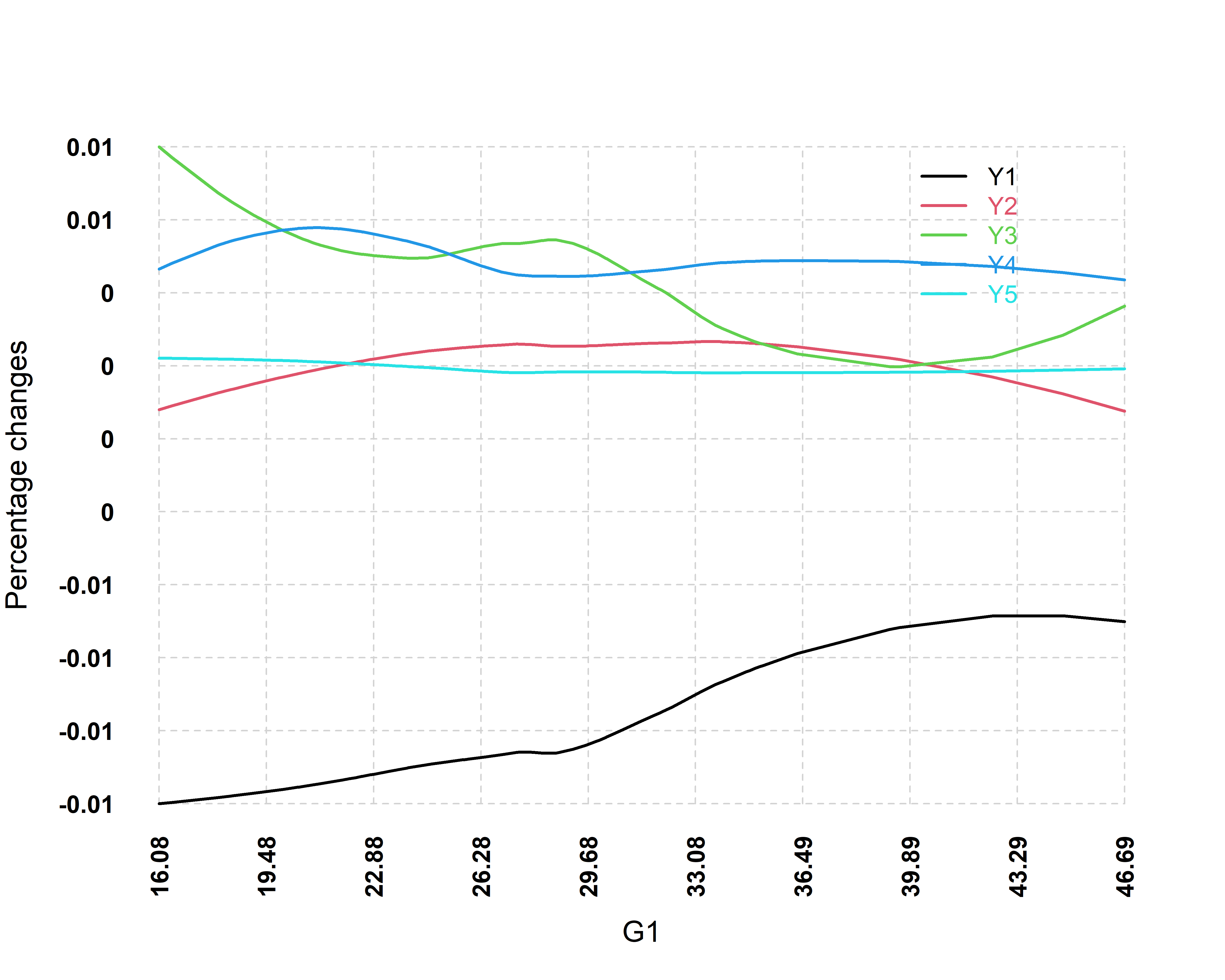}  \\
(a) ICE plot for HII  &  (b) MEs for HII \\
\includegraphics[scale = 0.3]{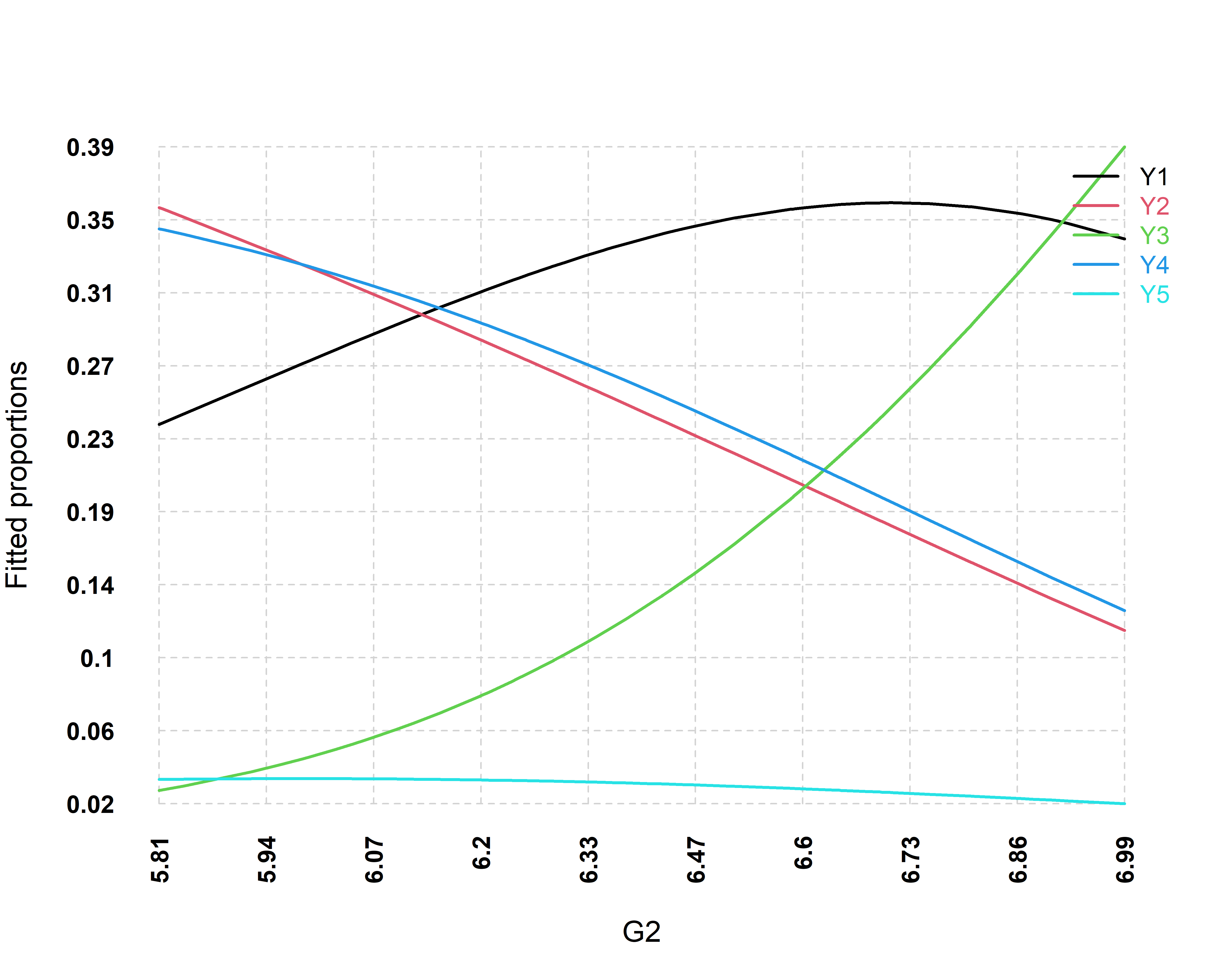}  &
\includegraphics[scale = 0.3]{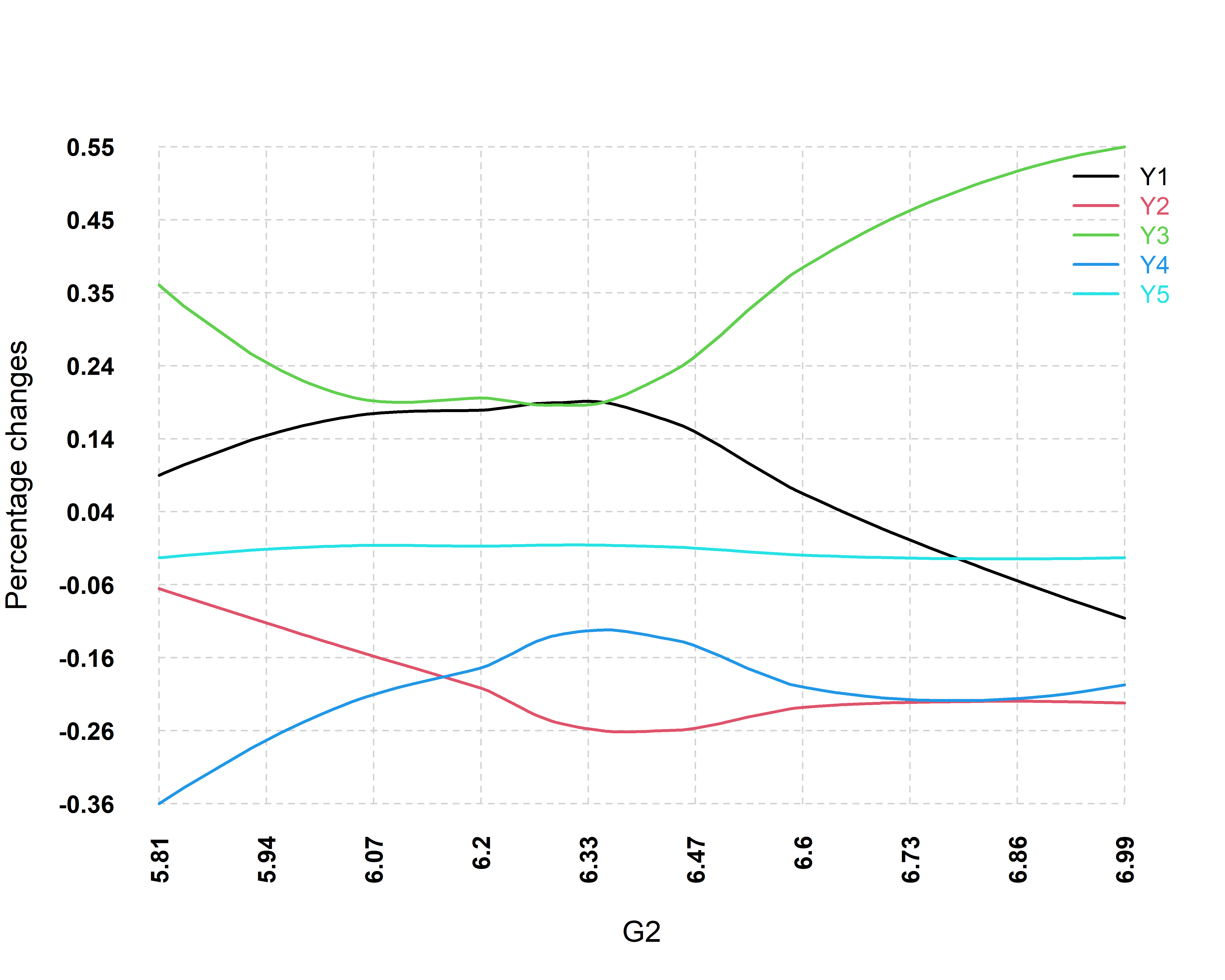}  \\
(c) ICE plot for CaCl$_2$ &  (d) MEs for CaCl$_2$ \\
\includegraphics[scale = 0.3]{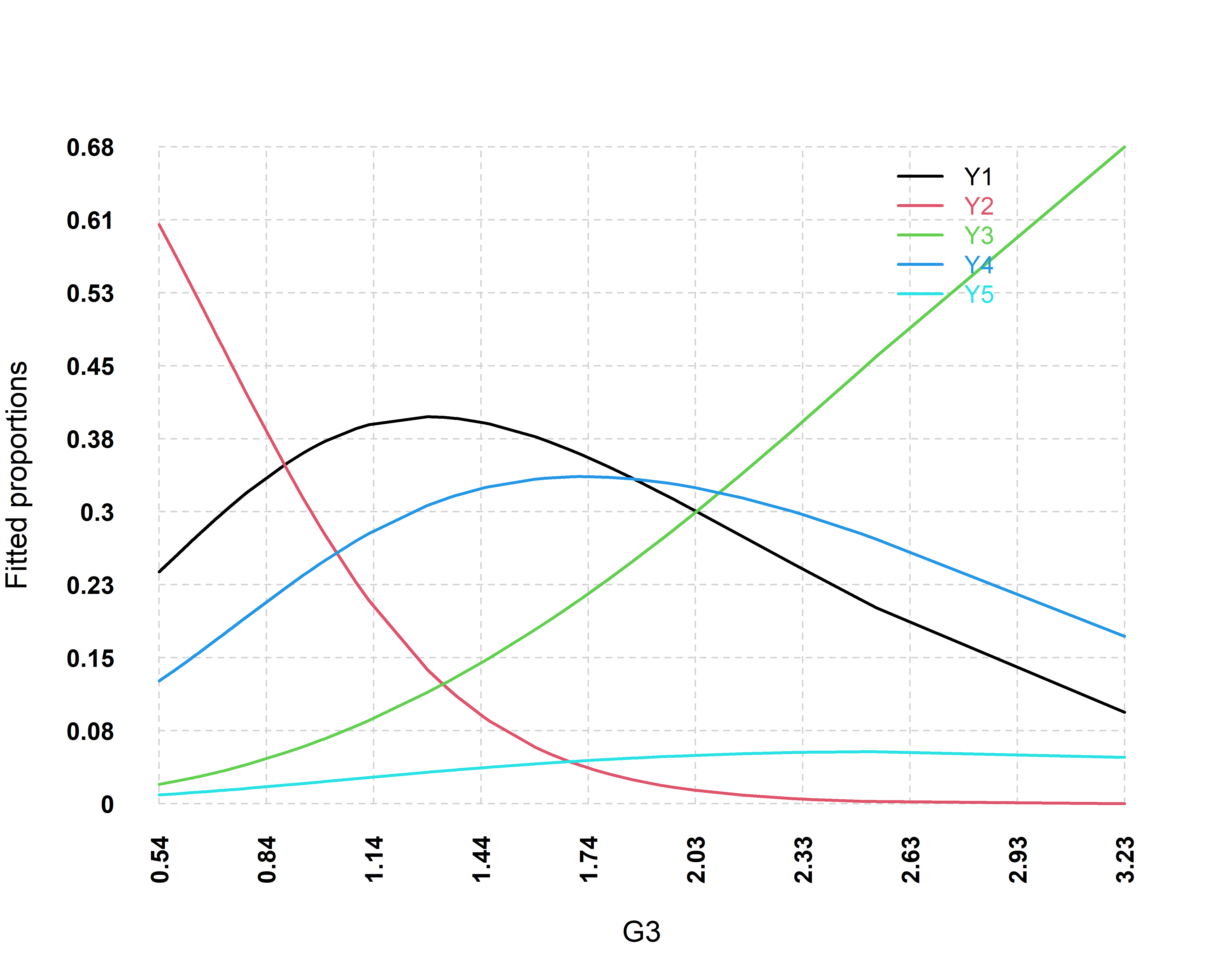}  &
\includegraphics[scale = 0.3]{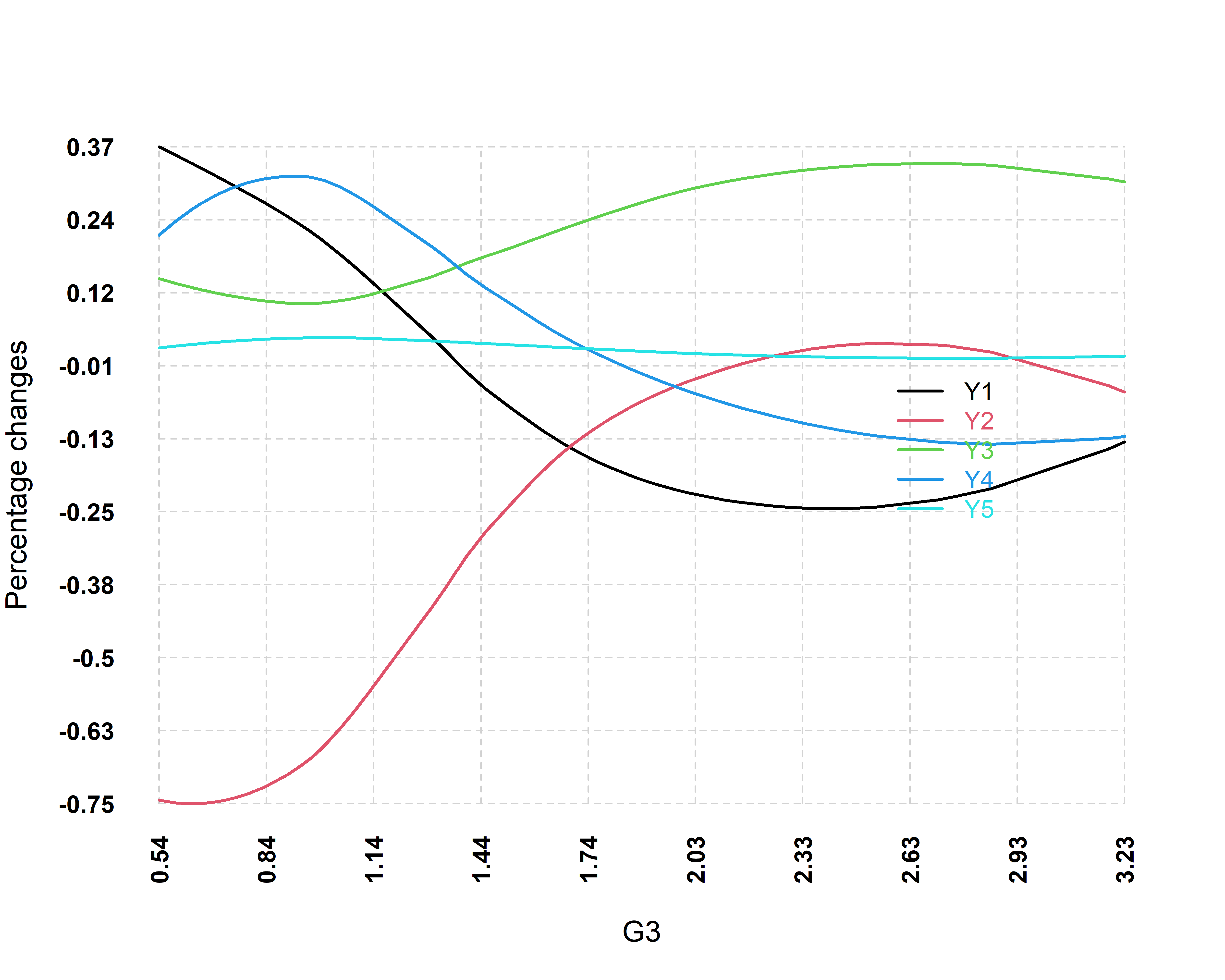}  \\
(e) ICE plot for SOC  &  (f) MEs for SOC  \\ 
\end{tabular}
\caption{Agricultural economics dataset: ICE and MEs plots for the three predictor variables. \label{ice.me} }
\end{figure}

As mentioned earlier, the interpretation of the estimated regression coefficients of the $\alpha$--regression (and of its spatial extensions) is rather limited. For this reason we may rely on ICE plots to visualize the effect of each predictor variable on the fitted compositions. Figure \ref{ice.me} contains the ICE (left column) and MEs\footnote{These are the smoothed changes after fitting a locally polynomial surface via the command \texttt{loess()} in \textit{R}.} (right column) plots for each of the three predictor variables of the Agricultural economics dataset, using $\alpha=1$. Inspection of the ICE plots reveals that the effect of HII is almost linear on the compositional responses, whereas the effects of CaCl$_2$ and SOC are not. Examination of the MEs plot shows the percentage--wise change at infinitesimal changes of each of the predictor variable. MEs play an important role in econometrics, but sometimes the estimated percentage change in regression models involving proportions yields high, and possibly unrealistic, values. In this example we see that even for small changes in CaCl$_2$ and SOC, the expected change in the compositional responses are significantly high. 

\subsubsection{Comparison of the \texorpdfstring{$\alpha$--regression}{alpha-regression}, without spatial dependence, to the \texorpdfstring{$\alpha$--$k$--$NN$}{alpha-k-NN} regression}
We will now compare the plain $\alpha$--regression to the $\alpha$--$k$--$NN$ regression for both datasets, ignoring the spatial information. The $\alpha$--$k$--$NN$ regression is a non--linear regression that extends the $k$--$NN$ to compositional data via the $\alpha$--transformation, which was shown to be on par with or to outperform the Kullback--Leibler divergence regression. We performed the 10--fold CV, repeated 10 times, and each time computed the KLD and the running time of the CV for each algorithm. We considered
21 values for $\alpha$, ranging from $-1$ up to $1$ at a step size of $0.1$. 

 Table \ref{comp} presents the results. Regarding the Agricultural economics dataset, the performance of the $\alpha$--regression is only 2\% worse, and it is nearly 9 times slower. Regarding the Meuse river dataset, the $\alpha$--regression outperformed the $\alpha$--$k$--$NN$ by 4 times, but it is 4.3 times slower. 

\begin{table}[ht]
\centering
\caption{Comparison of $\alpha$--regression to $\alpha$--$k$--$NN$ regression in terms of predictive performance (KLD) and running time (in seconds).}
\label{comp}
\begin{tabular}{r|rr|rr}
\toprule
& \multicolumn{2}{c}{Agricultural economics} & \multicolumn{2}{c}{Meuse river} \\
\midrule
Model & KLD & Running time & KLD & Running time \\ \midrule
$\alpha$--regression           & 0.642 &  0.860 & 0.006 & 0.284 \\ 
$\alpha$--$k$--$NN$ regression & 0.629 & 0.099  & 0.027 & 0.066 \\
\bottomrule
\end{tabular}
\end{table}

\subsubsection{Comparison of the $\alpha$--regression with compositional predictors to the $\alpha$--SCLS model}
The $\alpha$--SCLS model was devised for compositional regression with compositional predictors and it allows for the inclusion of the $\alpha$--transformation. Table \ref{res3} contains the average results for the KLD, the optimal $\alpha$ selected, and the running time (in seconds) of each of the regression models applied to the Greek national elections dataset. The $\alpha$--regression produced a predictive KLD that is more than 5 times lower than that of the $\alpha$--SCLS at the cost of being multiple times slower. Once again, the value of the optimal $\alpha$ is far from 0, providing evidence that again the value of $\alpha=0$, corresponding to the ilr transformation, is not the optimal choice. Note also, that even though the compositional predictor consists of 6 components, the optimal value of the reduced dimensionality was 3.

\begin{table}[ht]
\centering
\caption{Greek national elections dataset: average results regarding the optimal choice of $\alpha$, KLD and running time (in seconds) of the 10--fold CV protocol, for the two models.}
\label{res3}
\begin{tabular}{rrrr}
\toprule
Model                & KLD   & $\alpha$ & Running time \\ \midrule
$\alpha$--SCLS       & 0.016 & 1.000    & 0.039 \\ 
$\alpha$--regression & 0.003 & 0.740    & 7.235 \\ 
\bottomrule
\end{tabular}
\end{table}

\subsection{Temporal $\alpha$--regression}
The task of interest is to perform a time series analysis using the Catalan elections dataset. This dataset was used in \cite{tsagris2025} to assess the $\alpha$--SCLS model in the time series framework. The temporal $\alpha$--regression and the $\alpha$--SCLS models were fitted to the data from the years 1980 up to 2003, and the data from the 2006 year of elections were considered to be the test set. Table \ref{res4} presents the results for the predictive KLD value, the optimal value of $\alpha$ and the running time. Both models chose the value $\alpha=1$, the predictive performance of the $\alpha$--SCLS model was nearly 2.5 times higher than that of the $\alpha$--regression, but the latter was, computationally, more expensive.

\begin{table}[ht]
\centering
\caption{Catalan elections dataset: The results regarding the optimal choice of $\alpha$, KLD and running time (in seconds) for the two models.}
\label{res4}
\begin{tabular}{rrrr}
\toprule
Model                & KLD   & $\alpha$ & Running time \\ \midrule
$\alpha$--SCLS       & 0.066 & 1.000    & 0.08 \\ 
$\alpha$--regression & 0.027 & 1.000    & 6.62 \\ 
\bottomrule
\end{tabular}
\end{table}

\subsection{Spatial regression models}
We review these two datasets again, but including the spatial dependence, for which case the spatial 10--fold CV was employed to determine the values of the optimal hyper--parameters in each of the five regression models. To reduce the computational burden, 5 values for $\alpha$ were chosen, namely $\alpha=(0.1,0.25,0.5,0.75,1)$. The values of $k$ (for the $\alpha$--SLX model) were set to $k=(2,\ldots,15)$, and the bandwidth $h$ of the GW$\alpha$R was initially set equal to the median of the distances of the coordinates. A grid of 19 values spanning from $0.1h$ up to $10h$ was used for the GW$\alpha$R model. The spatial 10--fold CV\footnote{The spatial 10--fold CV was also applied to the $\alpha$--regression to ensure a fair comparison.} was repeated 10 times\footnote{The time required to create the spatial folds was not accounted for.} and the results were aggregated over these 10 times. 

\subsubsection{Agricultural economics}
\begin{figure}[h!]
\centering
\includegraphics[scale=0.65, trim = 0 50 0 20]{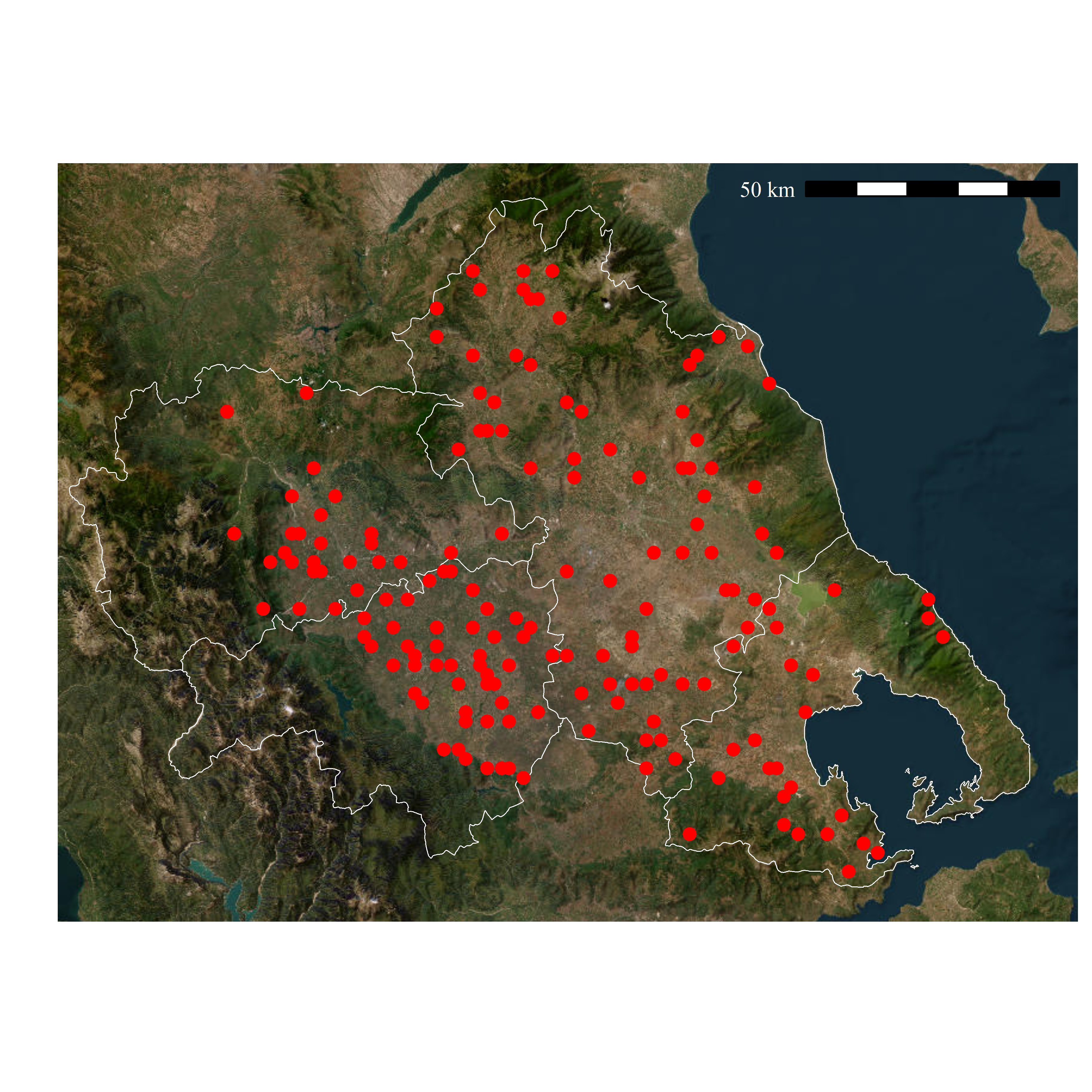} 
\caption{Region of Thessaly within Greece with the locations of the farms. \label{fig_real} }
\end{figure}

Figure \ref{fig_real} shows the Thessaly region in Greece, with the locations of the farms. The majority of the farms cultivate cereals and only a small proportion of farms cultivate grapes and wine. Specifically, 84.52\% of the farms cultivate cereals, 50.00\% cultivate cotton, 40.48\% maintain tree crops, 81.55\% hold other annual crops and pasture, and finally only 16.67\% of the farms cultivate grapes and wine. 

Table \ref{res} contains the results of the predictive performance estimation of the five models for the Agricultural economics, aggregated over the 10 times repeated spatial 10--fold CV protocol. The $\alpha$--SAR model exhibited the optimal predictive performance, and, unexpectedly, the use of spatially--lagged covariates deteriorated this performance. Two potential explanations that may account for this result are: (a) the particular dataset may not exhibit strong spatial spillover effects and (b) this suggests potential overfitting. In contrast, the GW$\alpha$R model was the computationally most expensive among them. The $\alpha$--ESF was the second best, with a predictive KLD slightly higher than that of the $\alpha$--SAR, but more than 120 times faster. 

\begin{table}[ht]
\centering
\caption{Agricultural economics dataset: average results regarding the optimal choice of $\alpha$, $k$, $h$, KLD and running time (in seconds) of the spatial 10--fold CV protocol, for each of the five models.}
\label{res}
\begin{tabular}{r|ccccr}
\toprule
Model                & KLD   & $\alpha$ & $k$ & $h$   & Running time \\  \midrule
$\alpha$--regression & 0.810 & 0.775    &     &       & 1.103   \\ 
$\alpha$--SLX        & 1.603 & 0.550    & 6   &       & 57.620  \\ 
$\alpha$--SAR        & 0.608 & 1.000    & 5   &       & 629.465 \\ 
GW$\alpha$R          & 0.869 & 0.675    &     & 3.369$\times 10^{-3}$ & 711.140 \\ 
$\alpha$--ESF        & 0.618 & 0.950    &     &       & 5.146   \\
\bottomrule
\end{tabular}
\end{table}

We then fitted the regression models using the optimal parameters obtained based on the CV protocol and computed the correlations (component-wise) between the observed and fitted compositions. Table \ref{cors} contains these correlations. GW$\alpha$R contains the highest correlations, with $\alpha$--ESF coming second. The $\alpha$--SAR model has achieved the third best fit. It is important to highlight that the spatial autoregressive parameter $\rho$ of the $\alpha$--SAR model was equal to $-0.148$ with a standard error equal to $0.0681$. A $t$--test indicates that the coefficient is statistically significant.  

\begin{table}[ht]
\centering
\caption{Agricultural economics dataset: Pearson correlations between each pair of the observed and fitted components for each of the five regression models.}
\label{cors}
\begin{tabular}{r|rrrrr}
\toprule
                     & Cereals & Cotton & Tree crops & Other annual crops & Grapes and wine\\   
                     &         &         &           & and pasture        &    \\ \midrule
$\alpha$--regression & 0.324 & 0.589 & 0.603 & 0.348 & 0.224\\ 
$\alpha$--SLX        & 0.327 & 0.626 & 0.627 & 0.413 & 0.328 \\ 
$\alpha$--SAR        & 0.318 & 0.582 & 0.608 & 0.357 & 0.230 \\
GW$\alpha$R          & 0.497 & 0.741 & 0.760 & 0.485 & 0.372 \\ 
$\alpha$--ESF        & 0.365 & 0.654 & 0.707 & 0.417 & 0.197 \\
\bottomrule
\end{tabular}
\end{table}

\begin{figure}[!ht]
\centering
\begin{tabular}{cc}
\includegraphics[scale = 0.3]{ice_G1.png}  &
\includegraphics[scale = 0.3]{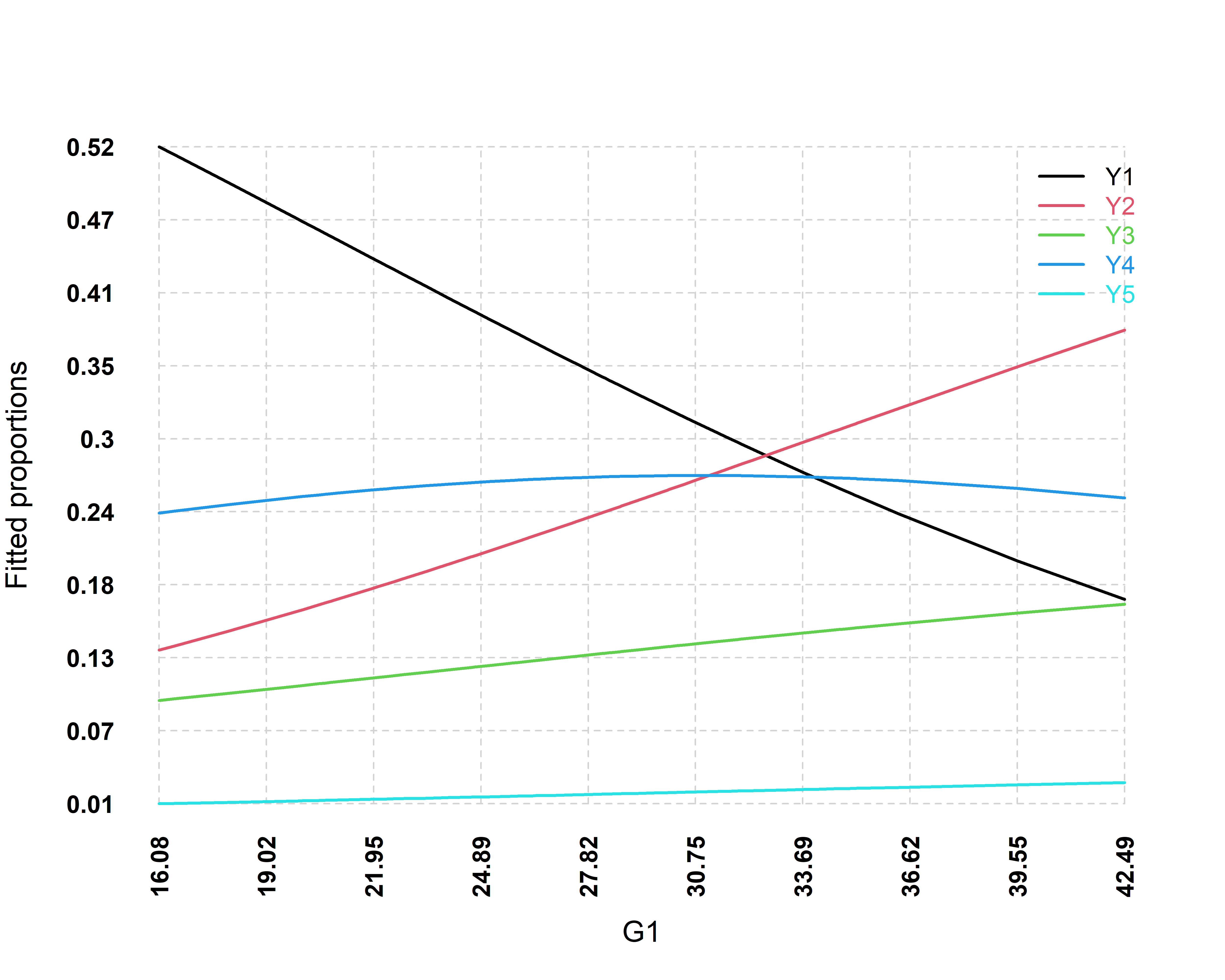}  \\
(a) $\alpha$--regression  &  (b) $\alpha$--ESF \\
\includegraphics[scale = 0.3]{ice_G2.png}  &
\includegraphics[scale = 0.3]{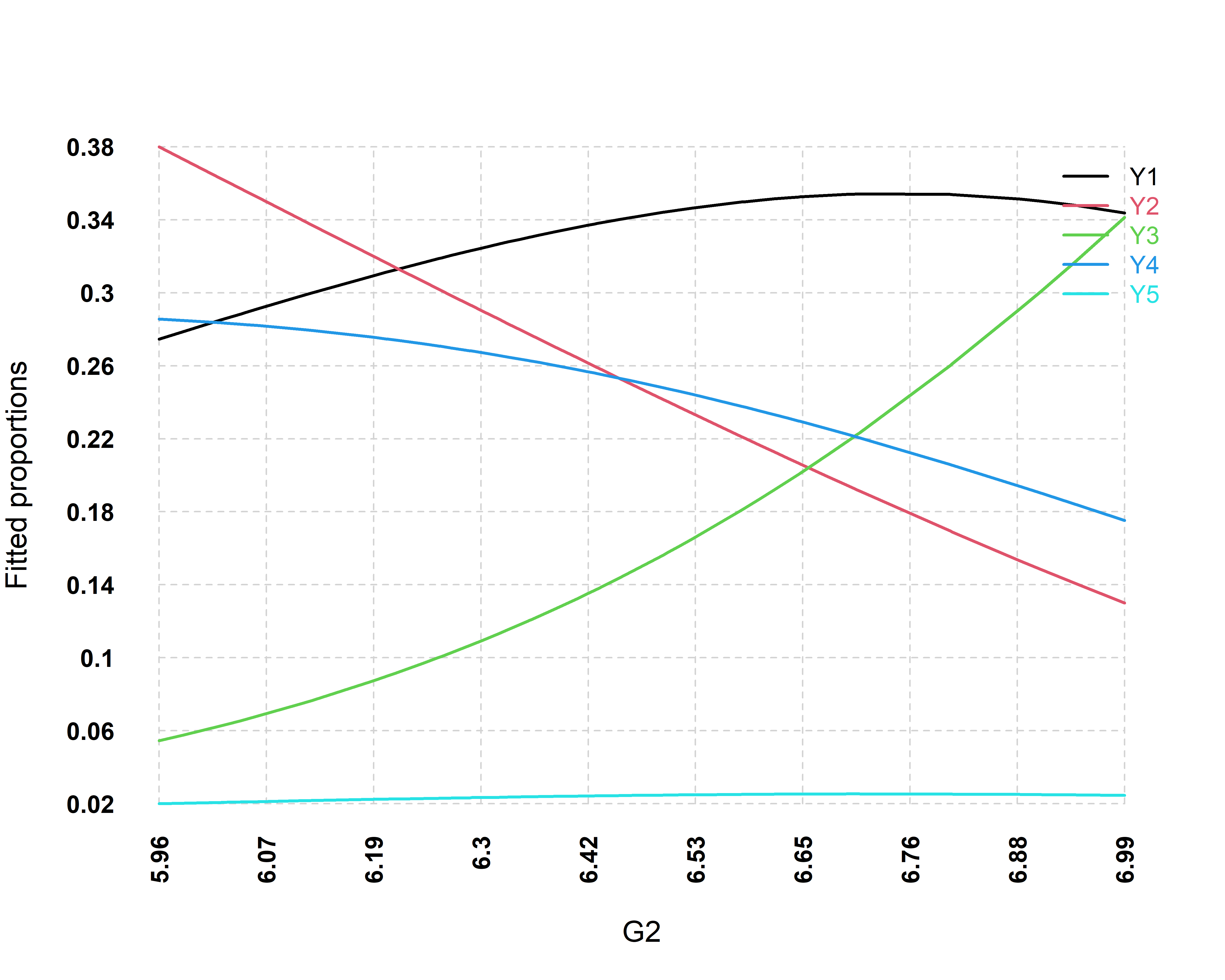}  \\
(c) $\alpha$--regression  &  (d) $\alpha$--ESF \\
\includegraphics[scale = 0.3]{ice_G3.png}  &
\includegraphics[scale = 0.3]{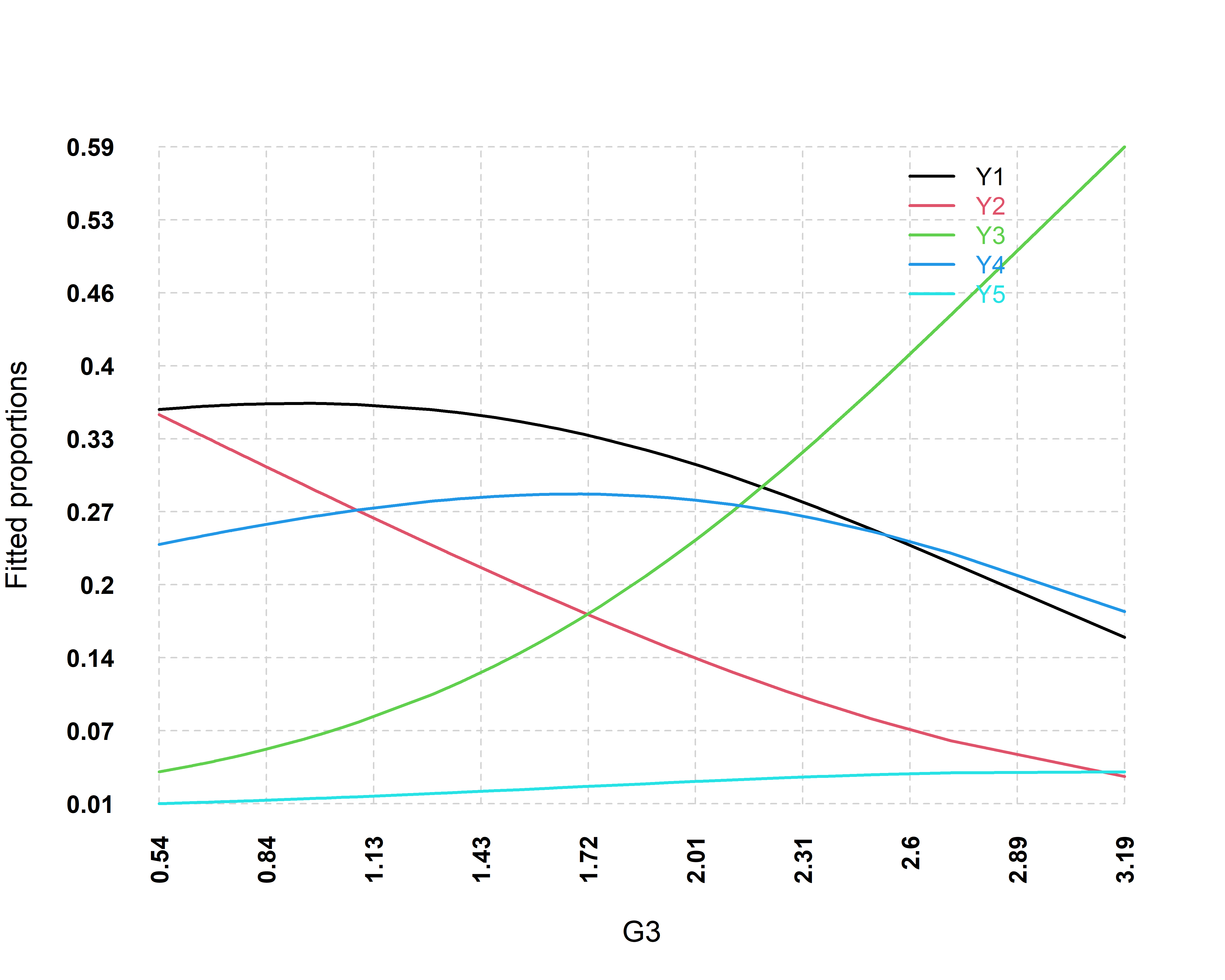}  \\
(e) $\alpha$--regression  &  (f) $\alpha$--ESF \\
\end{tabular}
\caption{Agricultural economics dataset: ICE plots for the three predictor variables using the $\alpha$--regression and the $\alpha$--ESF. (a)--(b): HII, (c)--(d): CaCl$_2$, (e)--(f): SOC. \label{ice.ice} }
\end{figure}

Figure \ref{ice.ice} contains the ICE plots for the three predictor variables. The first column is the same as that of Figure \ref{ice.me}, but we present it again here for ease of comparison between the (non--spatial) $\alpha$--regression and its spatial extension ($\alpha$--ESF). Notable differences are observed in Figures (a) and (b), for the HII. Other annual crops and pasture (Y4) increases according to $\alpha$--regression, but according to the $\alpha$--ESF model it increases and then decreases. Cotton (Y2) on the other hand, increases slightly based on $\alpha$--regression, whereas according to the $\alpha$--ESF model it increases monotonically. Regarding the third predictor variable (SOC), the overall trends are similar, but the curves are different.

\subsubsection{Meuse river dataset}
\begin{figure}[h!]
\centering
\includegraphics[scale=0.65]{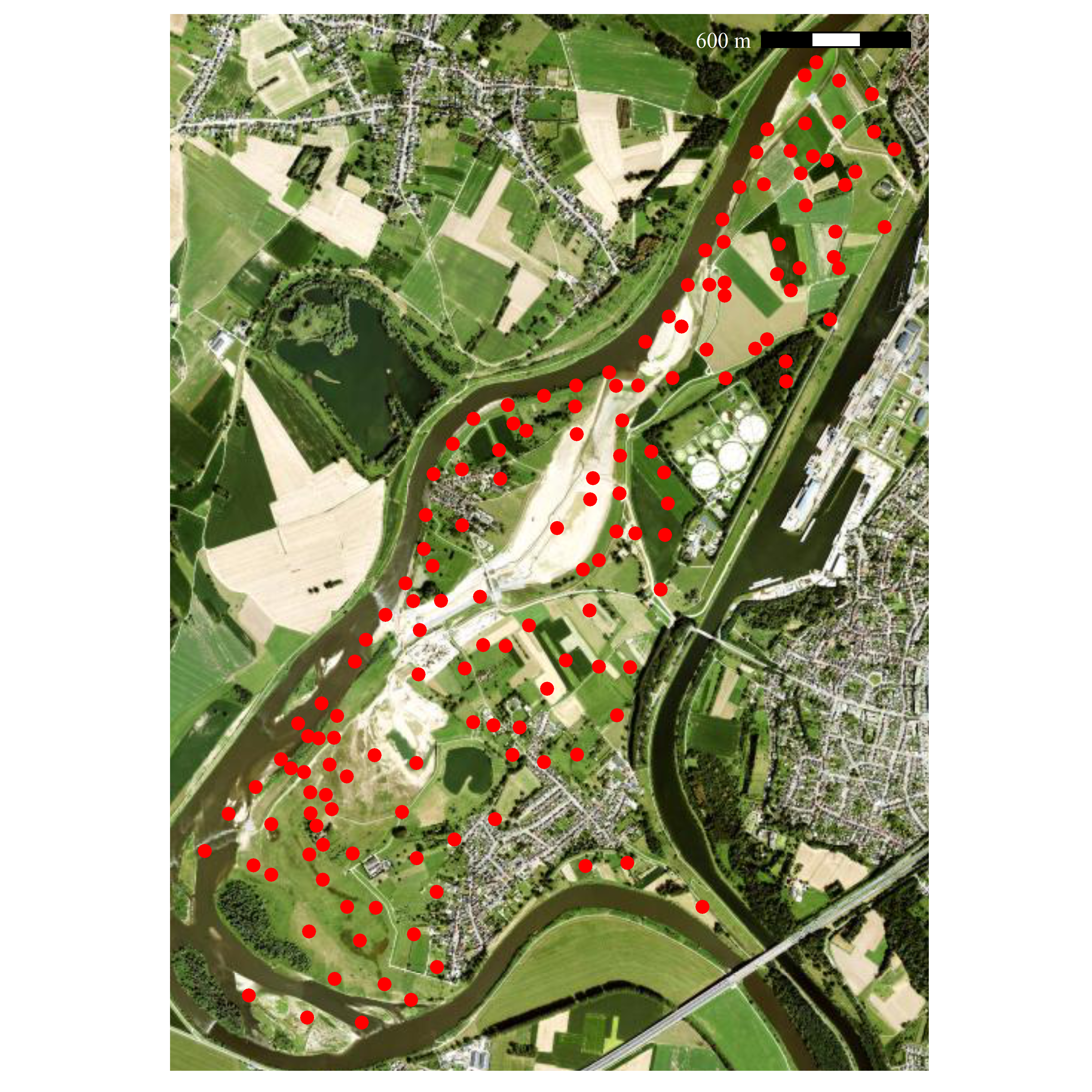}   
\caption{The flood plain of the river Meuse in the Netherlands. \label{fig_real2} }
\end{figure}

Figure \ref{fig_real2} shows the map with locations of the Meuse river dataset. This dataset is characterized by the absence of zero values. Table \ref{res2} contains the average results regarding the optimal choice of $\alpha$, $k$, $h$, KLD and running time (in seconds) for each of the five models for the Meuse river. In this case the $\alpha$--regression exhibits  performance comparable to the $\alpha$--SLX model, and the GW$\alpha$R model exhibited the optimal performance, at the cost of duration. Moreover, the optimal $\alpha$ value did not remain consistent across models, and the value of zero, that corresponds to the ilr transformation (\ref{ilr}), was never selected. 

\begin{table}[ht]
\centering
\caption{Meuse river dataset: average results regarding the optimal choice of $\alpha$, $k$, $h$, KLD and running time (in seconds) of the spatial 10--fold CV protocol, for each of the five models.}
\label{res2}
\begin{tabular}{r|ccccr}
\toprule
Model                & KLD   & $\alpha$ & $k$ & $h$   & Running time \\ \midrule
$\alpha$--regression & 0.006 & 0.500    &     &       & 0.339  \\
$\alpha$--SLX        & 0.006 & 0.500    & 3   &       & 9.956   \\ 
$\alpha$--SAR        & 0.006 & 0.435    & 4   &       & 185.362 \\   
GW$\alpha$R          & 0.036 & 0.250    &     & 590.523$\times 10^{-6}$ & 276.368 \\ 
$\alpha$--ESF        & 0.006 & 0.550    &     &       & 22.437  \\ 
\bottomrule
\end{tabular}
\end{table}
Table \ref{cors2} contains the correlations between each pair of the observed and fitted components for each of the five regression models. The $\alpha$--ESF regression model outperformed the rest, with the $\alpha$--SLX coming second. The $\alpha$--ESF model has the highest correlations, while the GW$\alpha$R model has received the third place. It is important to highlight that the spatial autoregressive parameter $\rho$ of the $\alpha$--SAR model was equal to $-0.206$ with a standard error equal to $0.164$.

\begin{table}[ht]
\centering
\caption{Meuse river dataset: Pearson correlations between each pair of the observed and fitted components for each of the five regression models.}
\label{cors2}
\begin{tabular}{r|rrrr}
\toprule
                     & Cadmium & Copper & Lead & Zinc \\   \midrule
$\alpha$--regression & 0.638 & 0.543 & 0.471 & 0.628 \\ 
$\alpha$--SLX        & 0.717 & 0.575 & 0.506 & 0.653 \\ 
$\alpha$--SAR        & 0.592 & 0.559 & 0.472 & 0.634 \\
GW$\alpha$R          & 0.648 & 0.558 & 0.482 & 0.638 \\ 
$\alpha$--ESF        & 0.818 & 0.800 & 0.769 & 0.818 \\
\bottomrule
\end{tabular}
\end{table}

\subsection{Effect of the $\alpha$--transformation}
\begin{figure}[!ht]
\centering
\begin{tabular}{cc}
\includegraphics[scale = 0.3]{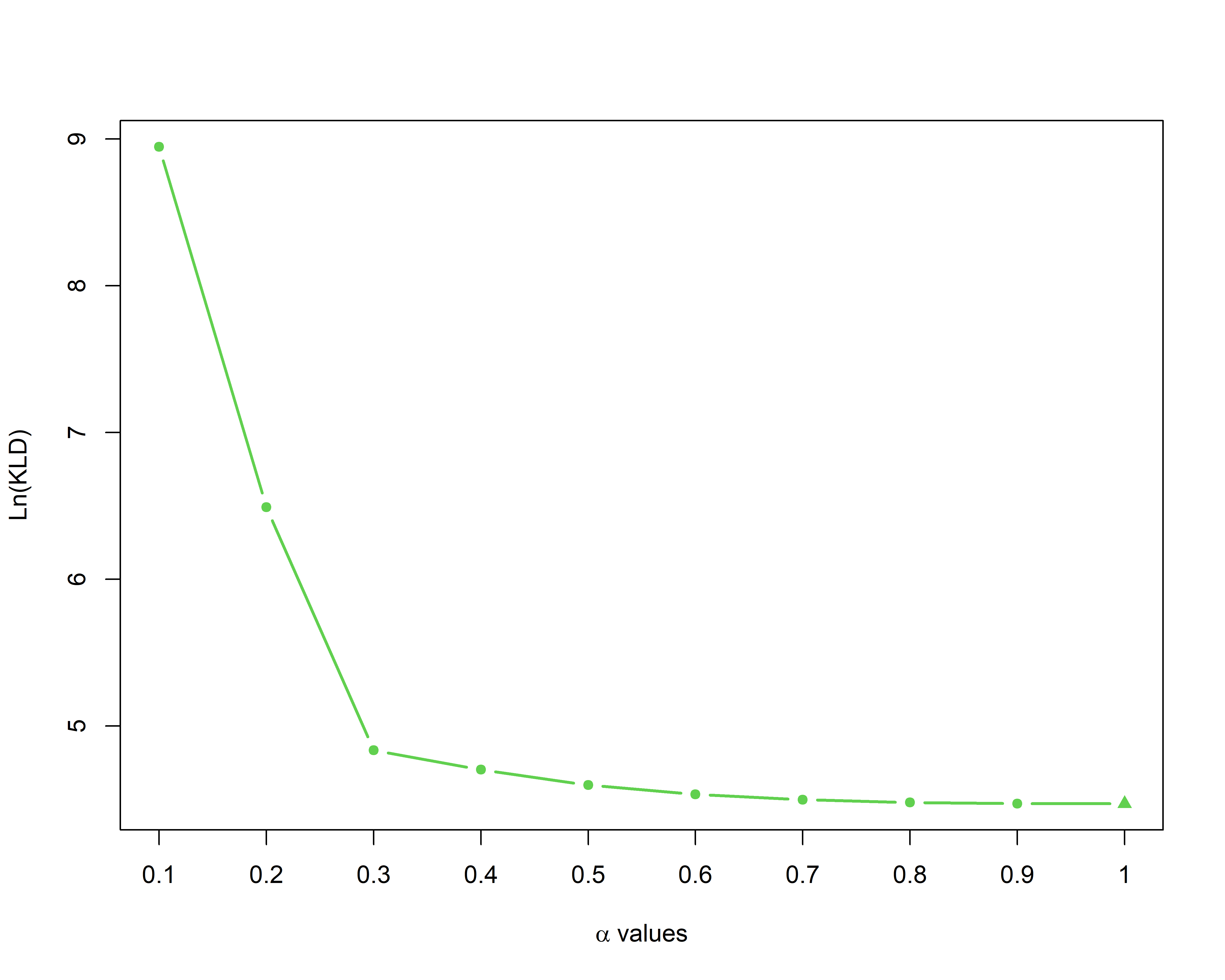}  &
\includegraphics[scale = 0.3]{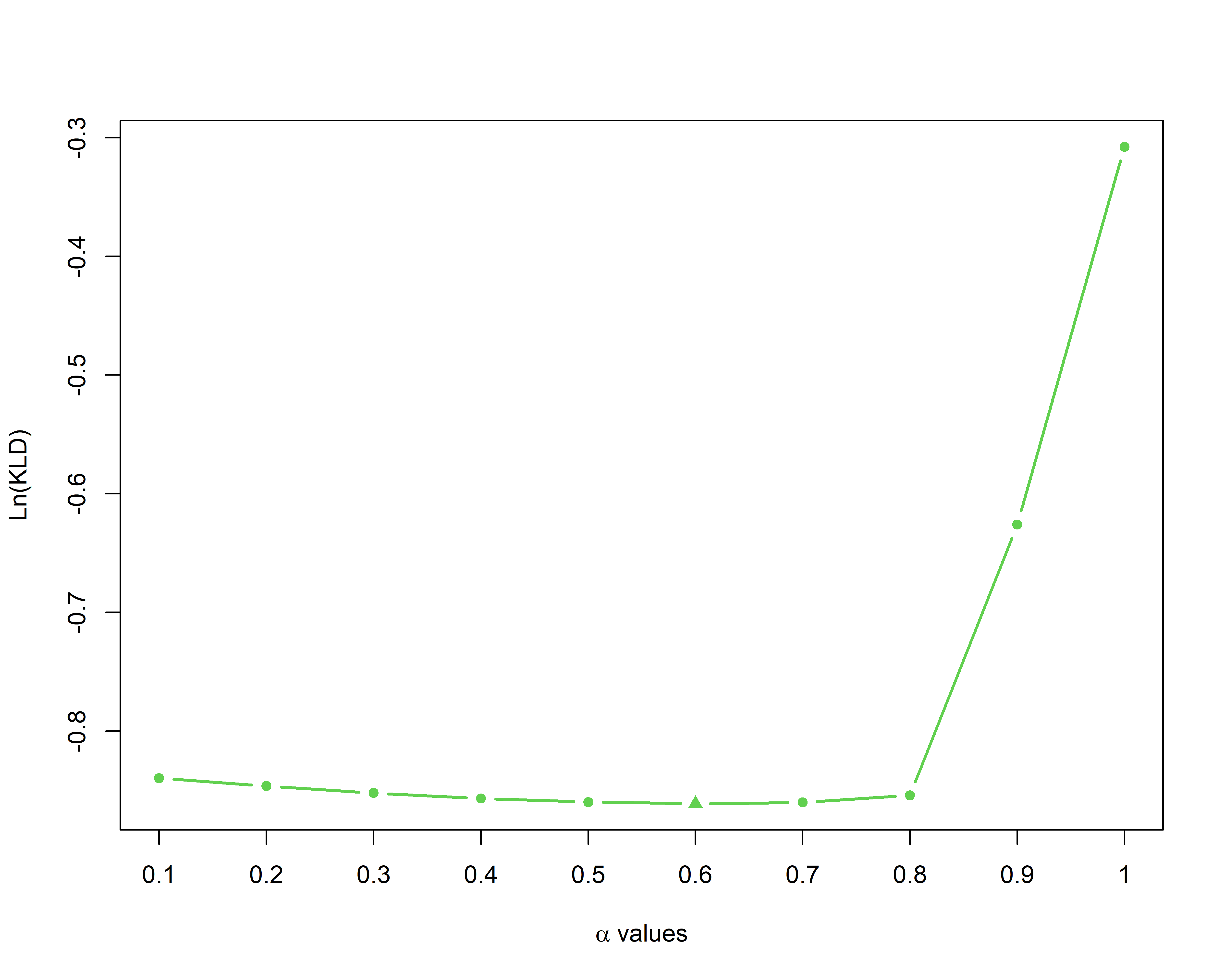}  \\
(a) Agricultural economics  &  (b) Meuse river \\
\end{tabular}
\caption{KLD values (in log--scale) versus the values of $\alpha$ of the $\alpha$--ESF model applied to the Agricultural economics and the Meuse river datasets. The triangle denotes the value of $\alpha$ that minimizes the KLD value. \label{aklds} }
\end{figure}

We mentioned in Section \ref{sec::limit} that the limiting case of $\alpha \rightarrow 0$ yields the ilr transformation (\ref{ilr}). This implies that with zero values present choosing a small value of $\alpha$ would approximate the results of the LRA approach. We performed a sensitivity analysis on the choice of $\alpha$. We applied the $\alpha$--ESF model for a range of values of $\alpha$ and each time computed the KLD using the fitted values. Figure \ref{aklds} presents the results for (a) the Agricultural and (b) the Meuse river datasets. For the Agricultural economics dataset, the minimum KLD value is observed at $\alpha=1$, whereas for the Meuse river dataset at $\alpha=0.6$. Both values are in close agreement with the spatial 10--fold CV protocol. 

\clearpage
\section{Conclusions} \label{sec::concl}
The $\alpha$--regression \citep{tsagris2015b} was revisited, and its theoretical properties were rigorously examined. We further proposed a permutation--based test to inspect the statistical significance of the estimated regression coefficients. Their interpretation is not straightforward and we computed the MEs, and employed ICE plots to visualize the effect of each predictor on the fitted compositions. Inspection of the ICE plots showed that the regression coefficients are of limited use in interpreting the effect of the predictor variables. A robustified version based on quantile regression was proposed, although it is computationally intensive and may exhibit convergence difficulties. However, other robust versions were mentioned, which are still computationally heavier than the plain $\alpha$--regression. We also discussed how to include natural splines in this framework. The inclusion of compositional predictors was facilitated via PCA, but this makes the regression computationally more challenging as it involves a second $\alpha$ parameter and also the choice of the appropriate number of principal components to use. To alleviate the computational cost we used the same value of $\alpha$ on both sides. This PCA--based formulation further enabled the use of the $\alpha$--regression to model temporal associations.

The $\alpha$--regression was next expanded to account for spatial dependencies by introducing the $\alpha$--SLX, the $\alpha$--SAR, the GW$\alpha$R and the $\alpha$--ESF models. For all models, formulas for the MEs were provided and their capabilities were tested on two real datasets. ICE plots may still be used but only for the $\alpha$--SLX and the $\alpha$--ESF models. The $\alpha$--SAR depends on the response values of the neighbours and hence prohibits the use of ICE plots, while the GW$\alpha$R model already produces location--specific regression coefficients. The last proposed spatial model was the $\alpha$--ESF that is computationally more efficient. Within this framework we may use the previously mentioned permutation--based test can be used to test for spatial dependence. Evidently, the robust extensions of the $\alpha$--regression, the use of compositional predictors via PCA, and the temporal $\alpha$--regression formulation may be used with the spatial models as well. Among the four spatial models we propose the use of the $\alpha$--ESF model as it is a generic, computationally scalable and easily customizable model. A robustified version is applicable, and one can incorporate spatial spillovers, spatial dependence, splines, and compositional predictors in a straightforward manner. 

Using real data, the $\alpha$--regression model was shown to be on par or outperformed existing regression models. The empirical results indicated that the proposed $\alpha$--regression models perform competitively with, and in several cases outperform, existing approaches, particularly when zero components and spatial dependence are present. These findings suggest that LRA is not always optimal and that data--driven transformations offer a flexible alternative for compositional regression modelling. Compared to the $\alpha$--SCLS model \citep{tsagris2025}, the $\alpha$--regression has the flexibility to become highly non--linear using splines. Further, unlike previous models that were devised for specific cases, the $\alpha$--regression can be generalized to include compositional predictors and temporal dependence in a straightforward manner.  

Overall, the paper unified many regression models (such as models to include spatio--temporal dependencies and compositional predictors) for compositional data and generalized the ilr--based regression models. The \textit{R} package \textsf{CompositionalSR} \citep{compositionalsr2026} has been developed to perform all the $\alpha$--regression models, including functions to perform (spatial) K--fold CV, computation of the MEs and creation of ICE plots for the $\alpha$--regression, and is available to download from \href{https://cran.r-project.org/web/packages/CompositionalSR/index.html}{CRAN}. 

Future research could explore nonparametric spatially varying models for compositional data, as well as hybrid approaches that blend GWR with machine learning techniques for complex compositional systems. For example, the adoption of the $\alpha$--regression or of the $\alpha$--transformation within neural network frameworks is an interesting direction. An unexplored direction that is worth exploring is variable selection using the $\alpha$--regression. 

\bibliographystyle{apalike}
\bibliography{vivlio}

\section*{Appendix}
\begin{appendix}
\renewcommand{\theequation}{A.\arabic{equation}}
\setcounter{equation}{0}  

\section{Proof of Theorem~\ref{theorem:consistency}}\label{app:proof_consistency}

\begin{proof}
The proof follows the classical M--estimator consistency argument \citep[Theorem~5.7]{vaart1998asymptotic}; see also \citet[Theorem~2.1]{newey1994large}, \citet[Theorem~4.1.1]{amemiya1985advanced}, and \citet[Section~3.2]{gallant1987nonlinear} for the NLLS specialization. We proceed in three steps.

\medskip
\noindent\textbf{Step 1: Uniform convergence of $Q_n$ to $Q$.}

Define $f_j(\theta) = \norm{\bm{y}_{j,\alpha} - \bm{\mu}_{j,\alpha}(\theta)}^2$. We want to show that
\[
  \sup_{\theta \in \Theta} |Q_n(\theta) - Q(\theta)| \convp 0.
\]
By the parallelogram--law bound,
\[
  f_j(\theta)
  = \norm{\bm{y}_{j,\alpha} - \bm{\mu}_{j,\alpha}(\theta)}^2
  \leq 2\norm{\bm{y}_{j,\alpha}}^2 + 2\norm{\bm{\mu}_{j,\alpha}(\theta)}^2
  \leq 2\norm{\bm{y}_{j,\alpha}}^2 + 2\sup_{\theta' \in \Theta}\norm{\bm{\mu}_{j,\alpha}(\theta')}^2
  =: M_j.
\]
By Assumption~\ref{ass:continuity}, $\E[M_j] < \infty$. Since $\theta \mapsto f_j(\theta)$ is continuous on the compact $\Theta$ (Assumptions~\ref{ass:compact} and~\ref{ass:continuity}), and the class $\{f_j(\theta):\theta\in\Theta\}$ is dominated by the integrable envelope $M_j$, the Uniform Strong Law of Large Numbers (USLLN; \citealt[Theorem~2.1]{newey1994large}) yields
\[
  \sup_{\theta \in \Theta}\left|\frac{1}{n}\sum_{j=1}^n f_j(\theta) - \E[f_j(\theta)]\right|
  \xrightarrow{a.s.} 0,
\]
and in particular convergence holds in probability.

\medskip
\noindent\textbf{Step 2: $\thetao$ is the unique minimizer of $Q$.}

By Remark~\ref{rem:dgp_consistency}, correct specification (Assumption~\ref{ass:dgp}) implies $Q(\theta) \geq Q(\thetao)$ for all $\theta \in \Theta$. Assumption~\ref{ass:identification} strengthens this to: for every $\epsilon > 0$ there exists $\delta > 0$ such that $Q(\theta) \geq Q(\thetao) + \delta$ whenever $\norm{\theta - \thetao} \geq \epsilon$.

\medskip
\noindent\textbf{Step 3: Convergence of the minimizer.}

Fix $\epsilon > 0$ and let $\delta > 0$ be as in Assumption~\ref{ass:identification}. Since $\thetahat$ minimizes $Q_n$, we have $Q_n(\thetahat) \leq Q_n(\thetao)$. For any event on which $\norm{\thetahat - \thetao} \geq \epsilon$:
\begin{align*}
  Q(\thetao) + \delta
  &\leq Q(\thetahat) \\
  &= Q_n(\thetahat) + [Q(\thetahat) - Q_n(\thetahat)] \\
  &\leq Q_n(\thetao) + \sup_{\theta\in\Theta}|Q_n(\theta)-Q(\theta)| \\
  &= Q(\thetao) + [Q_n(\thetao) - Q(\thetao)] + \sup_{\theta\in\Theta}|Q_n(\theta)-Q(\theta)| \\
  &\leq Q(\thetao) + 2\sup_{\theta\in\Theta}|Q_n(\theta)-Q(\theta)|.
\end{align*}
Rearranging, $\delta \leq 2\sup_{\theta\in\Theta}|Q_n(\theta)-Q(\theta)|$. Therefore,
\[
  P\!\left(\norm{\thetahat - \thetao} \geq \epsilon\right)
  \leq P\!\left(\sup_{\theta\in\Theta}|Q_n(\theta)-Q(\theta)| \geq \delta/2\right)
  \to 0,
\]
where the last step uses Step~1. Since $\epsilon > 0$ was arbitrary, $\thetahat \convp \thetao$.
\end{proof}

\begin{remark}[Relationship to classical M--estimator theory]
The proof above is an instance of the general consistency theorem for M--estimators \citep[Theorem~5.7]{vaart1998asymptotic}. The key structural requirements---compactness of $\Theta$, continuity of the criterion, an integrable envelope, and a uniquely separated population minimum---are standard in the NLLS literature \citep[Theorem~8.1]{wooldridge2010econometric}. The present setting is non--standard in that the transformed response $\bm{y}_{j,\alpha}$ takes values in a bounded subset of $\R^{D-1}$ that depends on $\alpha$, but this does not affect the proof structure.
\end{remark}

\section{Proof of Theorem~\ref{theorem:normality}}\label{app:proof_normality}

\begin{proof}
The proof is a standard NLLS asymptotic normality argument; see \citet[Theorem~5.21]{vaart1998asymptotic}, \citet[Theorem~3.1]{newey1994large}, \citet[Theorem~4.1.3]{amemiya1985advanced}, and \citet[Theorem~8.3]{wooldridge2010econometric}. We proceed in three steps.

\medskip
\noindent\textbf{Step 1: First--order condition and Taylor expansion.}

Since $\thetao \in \operatorname{int}(\Theta)$ (Assumption~\ref{ass:compact}) and $\thetahat \convp \thetao$ (Theorem~\ref{theorem:consistency}), for all large enough $n$ the minimizer $\thetahat$ lies in $\operatorname{int}(\Theta)$ with probability approaching one. The smoothness assumption (Assumption~\ref{ass:smoothness}) then ensures that $\thetahat$ satisfies the first--order condition exactly:
\[
  \nabla_\theta Q_n(\thetahat) = \frac{2}{n}\sum_{j=1}^n g_j(\thetahat)^\top \bm{r}_j(\thetahat) = \bm{0}.
\]
Applying a mean value expansion about $\thetao$ to each component of $\nabla_\theta Q_n$, there exists a (random) $\bar\theta_n$ on the segment between $\thetahat$ and $\thetao$ such that
\[
  \bm{0}
  = \nabla_\theta Q_n(\thetao)
    + \nabla^2_\theta Q_n(\bar\theta_n)(\thetahat - \thetao).
\]
Rearranging and multiplying by $\sqrt{n}$:
\begin{equation}\label{eq:taylor}
  \sqrt{n}(\thetahat - \thetao)
  = -\left[\nabla^2_\theta Q_n(\bar\theta_n)\right]^{-1}
    \sqrt{n}\,\nabla_\theta Q_n(\thetao),
\end{equation}
provided $\nabla^2_\theta Q_n(\bar\theta_n)$ is invertible for large $n$ (established in Step~3).

\medskip
\noindent\textbf{Step 2: Central Limit Theorem for the score.}

The gradient of the sample criterion at $\thetao$ is
\[
  \nabla_\theta Q_n(\thetao)
  = \frac{2}{n}\sum_{j=1}^n g_j(\thetao)^\top \bm{r}_j(\thetao).
\]
We apply the multivariate CLT \citep[Theorem~2.7]{billingsley1999convergence} to the i.i.d.\ summands $\bm{s}_j := g_j(\thetao)^\top \bm{r}_j(\thetao)$. The mean is
\[
  \E[\bm{s}_j] = \E\!\left[g_j(\thetao)^\top \E[\bm{r}_j(\thetao)\mid\bm{x}_j]\right] = \bm{0},
\]
because $\E[\bm{r}_j(\thetao)\mid\bm{x}_j] = \E[\bm{y}_{j,\alpha}\mid\bm{x}_j] - \bm{\mu}_{j,\alpha}(\thetao) = \bm{0}$ by Assumption~\ref{ass:dgp}. The covariance matrix is
\[
  \E[\bm{s}_j \bm{s}_j^\top]
  = \E\!\left[g_j(\thetao)^\top \bm{r}_j(\thetao)\,\bm{r}_j(\thetao)^\top g_j(\thetao)\right]
  = \OmegaMat,
\]
which is finite by Assumptions~\ref{ass:moments} and~\ref{ass:continuity}. Therefore, the CLT gives
\[
  \frac{\sqrt{n}}{2}\,\nabla_\theta Q_n(\thetao)
  = \frac{1}{\sqrt{n}}\sum_{j=1}^n \bm{s}_j
  \convd N\!\left(\bm{0},\, \OmegaMat\right).
\]

\medskip
\noindent\textbf{Step 3: Convergence of the Hessian.}

We show that $\nabla^2_\theta Q_n(\bar\theta_n) \convp 2\GMat$. The Hessian of the individual criterion is
\[
  \nabla^2_\theta f_j(\theta)
  = 2\,g_j(\theta)^\top g_j(\theta)
    - 2\sum_{k=1}^{D-1} r_{jk}(\theta)\,\nabla^2_\theta \mu_{jk,\alpha}(\theta),
\]
where $r_{jk}$ and $\mu_{jk,\alpha}$ denote the $k$--th components of $\bm{r}_j$ and $\bm{\mu}_{j,\alpha}$, respectively. Therefore,
\[
  \frac{1}{2}\nabla^2_\theta Q_n(\theta)
  = \frac{1}{n}\sum_{j=1}^n g_j(\theta)^\top g_j(\theta)
    - \frac{1}{n}\sum_{j=1}^n \sum_k r_{jk}(\theta)\,\nabla^2_\theta \mu_{jk,\alpha}(\theta).
\]
By the USLLN (applicable because Assumption~\ref{ass:moments} provides the required envelope for both terms), the first sum converges uniformly to $\E[g_j(\theta)^\top g_j(\theta)]$ and the second sum converges uniformly to $\E[\sum_k r_{jk}(\theta)\,\nabla^2_\theta \mu_{jk,\alpha}(\theta)]$. Evaluating at $\theta = \thetao$ and invoking $\E[\bm{r}_j(\thetao)\mid\bm{x}_j]=\bm{0}$ (Assumption~\ref{ass:dgp}), the second expectation equals $\bm{0}$, giving
\[
  \frac{1}{2}\nabla^2_\theta Q(\thetao)
  = \E\!\left[g_j(\thetao)^\top g_j(\thetao)\right] = \GMat.
\]
Since $\bar\theta_n$ lies between $\thetahat$ and $\thetao$, and $\thetahat\convp\thetao$, we have $\bar\theta_n\convp\thetao$. By the uniform convergence and continuity of the Hessian in $\theta$,
\[
  \frac{1}{2}\nabla^2_\theta Q_n(\bar\theta_n) \convp \GMat.
\]
Since $\GMat$ is positive definite (Assumption~\ref{ass:nonsingular}), for all large $n$ the matrix $\nabla^2_\theta Q_n(\bar\theta_n)$ is invertible with probability approaching one, and $\left[\tfrac{1}{2}\nabla^2_\theta Q_n(\bar\theta_n)\right]^{-1} \convp \GMat^{-1}$.

\medskip
\noindent\textbf{Combining the steps.}

Substituting Steps~2 and~3 into~\eqref{eq:taylor} and applying Slutsky's theorem \citep[Theorem~2.8]{billingsley1999convergence}:
\begin{align*}
  \sqrt{n}(\thetahat - \thetao)
  &= -\left[\tfrac{1}{2}\nabla^2_\theta Q_n(\bar\theta_n)\right]^{-1}
     \cdot \frac{1}{\sqrt{n}}\sum_{j=1}^n \bm{s}_j \\
  &\convd -\GMat^{-1} \cdot N(\bm{0}, \OmegaMat)
   = N\!\left(\bm{0},\,\GMat^{-1}\OmegaMat\GMat^{-1}\right). \qedhere
\end{align*}
\end{proof}

\section{Gradient vector and Hessian matrix for the \texorpdfstring{$\alpha$--regression}{alpha-regression}}
The least squares objective function is
$$\text{SSE}(\bm{Y}, \bm{X}; \alpha, \bm{B}) = -\frac{1}{2}\text{tr}[(\bm{y}_{\alpha} - \bm{\mu}_{\alpha})^\top(\bm{y}_{\alpha} - \bm{\mu}_{\alpha})],$$
where $\bm{y}_{\alpha}$ is the $\alpha$--transformed observed compositional data ($n \times d$ matrix), $\bm{\mu}_{\alpha}$ is the $\alpha$--transformed fitted compositional values ($n \times d$ matrix), $n$ is the number of observations, and $d = D - 1$ where $D$ is the number of components in the composition. 

The fitted compositional values come from the inverse alr transformation:
$$\mu_1 = \frac{1}{1+\sum_{j=1}^d e^{x^\top\beta_j}}, \quad \mu_i = \frac{e^{x^\top\beta_{i-1}}}{1+\sum_{j=1}^d e^{x^\top\beta_j}}, \quad i=2,\ldots,D.$$

\subsection{The \texorpdfstring{$\alpha$--transformation}{alpha-transformation}}

The $\alpha$--transformation consists of two steps:

Step 1: Power transformation
$$u_i = \frac{\mu_i^\alpha}{\sum_{j=1}^D \mu_j^\alpha}, \quad i=1,\ldots,D.$$

Step 2: Helmert transformation
$$z = \frac{1}{\alpha}H(Du - j_D),$$
where $H$ is the $d \times D$ Helmert sub--matrix and $j_D$ is a $D$--dimensional vector of ones.

\subsection{First derivatives (gradient)}

\subsubsection{Main gradient formula}
$$\frac{\partial l(\alpha)}{\partial \beta_k} = \text{tr}\left[(\bm{y}_{\alpha} - \bm{\mu}_{\alpha})^\top \frac{\partial \bm{\mu}_{\alpha}}{\partial \beta_k}\right].$$

\subsubsection{Expanded gradient formula}
$$\frac{\partial l(\alpha)}{\partial \beta_k} = \sum_{i=1}^n \sum_{m=1}^d \sum_{\ell=1}^D \sum_{p=1}^D r_{\alpha,im} \cdot \frac{D}{\alpha} H_{m\ell} \cdot \frac{\partial u_{i\ell}}{\partial \mu_{ip}} \cdot \frac{\partial \mu_{ip}}{\partial \beta_k} \cdot x_i,$$
where $r_{\alpha,im} = y_{\alpha,im} - m_{\alpha,im}$ are the residuals in $\alpha$--transformed space, $H_{m\ell}$ is the $(m,\ell)$ element of the Helmert sub--matrix, and $x_i$ is the covariate vector for observation $i$.

\subsubsection{Jacobian of power transformation}
$$\frac{\partial u_{i\ell}}{\partial \mu_{ip}} = \begin{cases}
\displaystyle\frac{\alpha \mu_{i\ell}^{\alpha-1}}{\sum_{j=1}^D \mu_{ij}^\alpha}\left(1 - \frac{\mu_{i\ell}^\alpha}{\sum_{j=1}^D \mu_{ij}^\alpha}\right) & \text{if } \ell = p \\ [3ex]
\displaystyle-\frac{\alpha \mu_{i\ell}^\alpha \mu_{ip}^{\alpha-1}}{(\sum_{j=1}^D \mu_{ij}^\alpha)^2} & \text{if } \ell \neq p
\end{cases}.$$
Let $T_i = \sum_{j=1}^D \mu_{ij}^\alpha$. In compact form:
$$\frac{\partial u_{i\ell}}{\partial \mu_{ip}} = \frac{\alpha \mu_{ip}^{\alpha-1}}{T_i}\left(\delta_{\ell p} - \frac{\mu_{i\ell}^\alpha}{T_i}\right),$$
where $\delta_{\ell p}$ is the Kronecker delta.

\subsubsection{Jacobian of multinomial logit}
Let $S_i = 1 + \sum_{j=1}^d e^{x_i^\top\beta_j}$.

$$\frac{\partial \mu_{ip}}{\partial \beta_k} = \begin{cases}
-\mu_{i1} \mu_{ik} x_i & \text{if } p = 1 \\ 
\mu_{ik}(1-\mu_{ik}) x_i & \text{if } p = k+1 \\ 
-\mu_{ip} \mu_{ik} x_i & \text{if } p \neq 1, p \neq k+1
\end{cases},$$
where $\mu_{ik} = \mu_{i,k+1}$ (the $(k+1)$--th component of the composition).

\subsubsection{Vectorized gradient formula}
$$\frac{\partial l(\alpha)}{\partial \beta_k} = X^\top w_k,$$
where the weight vector $w_k \in \mathbb{R}^n$ has elements:
$$w_{k,i} = \left\{r_{\alpha,i}^\top \cdot \frac{D}{\alpha}H \cdot J_u(i) \cdot J_\mu(i,k)\right\}.$$
\subsubsection*{Diagonal contribution}
$$w_{k,i}^{\text{diag}} = \sum_{\ell=1}^D r_{\alpha,i\ell} H_{\ell} J_{u,\text{diag}}(i,\ell) J_\mu(i,\ell,k)$$
where $J_{u,\text{diag}}(i,\ell) = \frac{\alpha \mu_{i\ell}^{\alpha-1}}{T_i}\left(1 - \frac{\mu_{i\ell}^\alpha}{T_i}\right)$.

\subsubsection*{Off--diagonal contribution}
$$w_{k,i}^{\text{off--diag}} = -\frac{\alpha}{T_i^2}\left[\left(\sum_{\ell=1}^D r_{\alpha,i\ell} H_{\ell} \mu_{i\ell}^\alpha\right)\left(\sum_{p=1}^D \mu_{ip}^{\alpha-1} J_\mu(i,p,k)\right) - \sum_{\ell=1}^D r_{\alpha,i\ell} H_{\ell} \mu_{i\ell}^\alpha \mu_{i\ell}^{\alpha-1} J_\mu(i,\ell,k)\right].$$
Total Weight:
$$w_{k,i} = w_{k,i}^{\text{diag}} + w_{k,i}^{\text{off--diag}}.$$

\section{Gradient vector and Hessian matrix for the \texorpdfstring{$\alpha$--SAR model}{alpha-SAR model}}
The $\alpha$--SAR model minimizes the sum of squared errors (SSE):
\begin{equation*}
\text{SSE}(\bm{Y}, \bm{X}; \alpha, \rho, \bm{B}) = \ell\left(\bm{\theta}\right) = \sum_{i=1}^{n} \|\bm{y}_{i,\alpha} - \bm{\mu}_{i,\alpha}\|^2 = \sum_{i=1}^{n} (\bm{y}_{i,\alpha} - \bm{\mu}_{i,\alpha})^\top (\bm{y}_{i,\alpha} - \bm{\mu}_{i,\alpha})
\end{equation*}
where $\bm{\theta} = (\vec(\bm{B})^\top, \rho)^\top$ contains all parameters.

\subsection{Model specification review}

The fitted compositional values are:
\begin{equation*}
\mu_i = 
\begin{cases}
\dfrac{1}{1 + \sum\limits_{j=1}^{d} e^{\tilde{\bm{x}}_i^\top\bm{\beta}_j}} & \text{for } i = 1\\[2ex]
\dfrac{e^{\tilde{\bm{x}}_i^\top\bm{\beta}_{i-1}}}{1 + \sum\limits_{j=1}^{d} e^{\tilde{\bm{x}}_i^\top\bm{\beta}_j}} & \text{for } i = 2, \ldots, D
\end{cases}
\end{equation*}
where:
\begin{equation*}
\tilde{\bm{x}}_i = [\bm{S}(\rho)^{-1}\bm{x}]_i = [(I_n - \rho W)^{-1}\bm{x}]_i
\end{equation*}

The Hessian matrix has the block structure:
\begin{equation*}
\bm{H} = 
\begin{bmatrix}
\bm{H}_{\bm{\beta}\bm{\beta}} & \bm{H}_{\bm{\beta}\rho}\\
\bm{H}_{\rho\bm{\beta}} & H_{\rho\rho}
\end{bmatrix},
\end{equation*}

where:
\begin{itemize}
\item $\bm{H}_{\bm{\beta}\bm{\beta}}$ is $(dp) \times (dp)$: derivatives w.r.t. regression coefficients
\item $\bm{H}_{\bm{\beta}\rho}$ is $(dp) \times 1$: mixed derivatives
\item $\bm{H}_{\rho\bm{\beta}} = \bm{H}_{\bm{\beta}\rho}^\top$ by symmetry
\item $H_{\rho\rho}$ is $1 \times 1$: second derivative w.r.t.\ $\rho$
\end{itemize}

\subsection{First derivatives (gradient)}

\subsubsection{Gradient with respect to \texorpdfstring{$\bm{\beta}_k$}{beta_k}}

\begin{equation*}
\frac{\partial \ell}{\partial \bm{\beta}_k} = \sum_{i=1}^{n} \sum_{j=1}^{d} r_{i,\alpha,j} \frac{\partial \mu_{i,\alpha,j}}{\partial \bm{\beta}_k},
\end{equation*}
where $r_{i,\alpha,j} = y_{i,\alpha,j} - \mu_{i,\alpha,j}$ are the residuals.

Using the chain rule:
\begin{equation*}
\frac{\partial \mu_{i,\alpha,j}}{\partial \bm{\beta}_k} = \frac{D}{\alpha} \sum_{\ell=1}^{D} \sum_{p=1}^{D} H_{j\ell} \frac{\partial u_{i\ell}}{\partial \mu_{ip}} \frac{\partial \mu_{ip}}{\partial \bm{\beta}_k},
\end{equation*}
where $H_{j\ell}$ is the $(j,\ell)$ element of the Helmert sub--matrix.

\subsubsection{Gradient with respect to \texorpdfstring{$\rho$}{rho}}
\begin{equation*}
\frac{\partial \ell}{\partial \rho} = \sum_{i=1}^{n} \sum_{j=1}^{d} r_{i,\alpha,j} \frac{\partial \mu_{i,\alpha,j}}{\partial \rho}.
\end{equation*}

Using the chain rule through $\tilde{\bm{x}}$:
\begin{equation*}
\frac{\partial \mu_{i,\alpha,j}}{\partial \rho} = \sum_{p=1}^{D} \frac{\partial \mu_{i,\alpha,j}}{\partial \mu_{ip}} \frac{\partial \mu_{ip}}{\partial \rho},
\end{equation*}
where:
\begin{equation*}
\frac{\partial \mu_{ip}}{\partial \rho} = \sum_{s=1}^{p} \frac{\partial \mu_{ip}}{\partial \tilde{x}_{is}} \frac{\partial \tilde{x}_{is}}{\partial \rho}.
\end{equation*}

\end{appendix}

\end{document}